\documentclass[12pt]{article}
\usepackage{amssymb}
\usepackage{amsmath}
\usepackage{amsmath}
\usepackage{amsthm}
\usepackage{amsfonts}
\usepackage{graphicx}
\usepackage{bbm}
\usepackage{mathrsfs}
\usepackage{enumerate}
\usepackage{dsfont}
\usepackage{natbib}
\usepackage{multirow}
\usepackage{dsfont}
\usepackage{color}%
\providecommand{\U}[1]{\protect\rule{.1in}{.1in}}

\numberwithin{equation}{section}
\newtheorem{theorem}{Theorem}

\newtheorem{remark}{Remark}

\newtheorem{assumption}{Assumption}
\newtheorem{lemma}{Lemma}
\theoremstyle{definition}

\DeclareMathOperator{\E}{\text{E}}

\begin{document}

\title{Quantile Treatment Effects in Regression Kink Designs\thanks{First arXiv date: March 15, 2017.\bigskip}}
\author{
Heng Chen\thanks{Heng Chen: chhe@bankofcanada.ca. Currency Department, Bank of Canada, 234 Wellington Street, Ottawa, ON, K1A 0G9, Canada.\bigskip}
\\
Bank of Canada
\and
Harold D. Chiang\thanks{Harold D. Chiang: harold.d.chiang@vanderbilt.edu. Department of Economics, Vanderbilt University, VU Station B \#351819, 2301 Vanderbilt Place, Nashville, TN 37235-1819, USA.\bigskip} 
\\
Vanderbilt
\and 
Yuya Sasaki\thanks{Yuya Sasaki: yuya.sasaki@vanderbilt.edu. Department of Economics, Vanderbilt University, VU Station B \#351819, 2301 Vanderbilt Place, Nashville, TN 37235-1819, USA.\bigskip} \thanks{We thank Yingying Dong, Robert Moffitt, and participants at New York Camp Econometrics XIII for useful comments. All the remaining errors are ours.}
\\
Vanderbilt
}
\date{ }
\maketitle

\begin{abstract}
\setlength{\baselineskip}{7.5mm} 
The literature on regression kink designs develops identification results for average effects of continuous treatments \citep{card_lee_pei_weber2015}, average effects of binary treatments \citep{dong2018}, and quantile-wise effects of continuous treatments \citep{chiang_sasaki2019}, but there has been no identification result for quantile-wise effects of binary treatments to date.
In this paper, we fill this void in the literature by providing an identification of quantile treatment effects in regression kink designs with binary treatment variables. 
For completeness, we also develop large sample theories for statistical inference and a practical guideline on estimation and inference.

\end{abstract}

\newpage
%%%%%%%%%%%%%%%%%%%%%%%%%%%%%%%%%%%%%%%%%%%%%%%%%%%%%%%%%%%%%%%%%%%%%%%%%%%%%%%%%%%%%%%%%%%%%%%

\section{Introduction}

Theories of identification in regression kink designs are advanced by a few papers in the recent literature. 
\citet{card_lee_pei_weber2015} propose identification of average effects of continuous treatments.
\citet{dong2018} proposes identification of average effects of binary treatments.
\citet{chiang_sasaki2019} propose identification of quantile-wise effects of continuous treatments. 
To date, no theory has been proposed for identification of quantile-wise effects of binary treatments in regression kink designs.
This paper aims to fill this void in the literature.

Specifically, in regression kink designs with binary treatments, we show that a local Wald ratio of derivatives of certain conditional expectation functions can be used to identify the conditional distribution functions of the potential outcomes given the event of local compliance. 
These conditional distribution functions can be used in turn to identify the quantile treatment effects given the event of local compliance. 
Our identification argument parallels that of \citet{frandsen_frolich_melly2012}, who show that a local Wald ratio of certain conditional expectation functions can be used to identify the conditional distribution functions of potential outcomes given the event of local compliance in the context of regression discontinuity designs.
Because of the lack of discontinuity in our context of regression kink designs, however, our identification result entails the limit case of the event of local compliance, which amounts to the subpopulation to which the marginal treatment effects \citep{bjorklund_moffitt1987,heckman_vytlacil1999,heckman_vytlacil2005} are relevant. 
This is analogous to, and provides a quantile counterpart of the identification result by \citet{dong2018}.

Our identifying formula takes a form of local Wald ratios of \textit{derivatives} of functions.
Such a form is related to the identifying formulas of several papers in the existing literature.
These papers include 
\citet{dong_lewbel2015} -- also see \citet{cerulli_dong_lewbel_poulsen2017} -- who use a local Wald ratio of derivatives of conditional expectation functions to identify the average effect of changing the threshold location in regression discontinuity designs,
\citet{card_lee_pei_weber2015} who use a local Wald ratio of derivatives of conditional expectation functions to identify average effects of continuous treatments in regression kink designs,
\citet{dong2018} who use a local Wald ratio of derivatives of conditional expectation functions to identify average effects of binary treatments in regression kink designs, and
\citet{chiang_sasaki2019} who use a local Wald ratio of derivatives of conditional quantile functions to identify quantile-wise effects of continuous treatments in regression kink designs.
Differently from each of these papers, we use the difference of left-inverses of two local Wald ratios of derivatives of conditional expectation functions to identify quantile-wise effects of binary treatments in regression kink designs.

While we motivate this paper by quantile treatment effects, the identifying formulas we provide as the main result of this paper can be also used to identity the distributional treatment effects.
Therefore, this paper also relates to \citet{abadie2002} who uses a form of Wald ratios to identify distributional treatment effects, and more closely relates to \citet{shen_zhang2016} who consider distributional treatment effects in the context of regression discontinuity designs.

In addition to the main identification result, we also provide methods of estimation and inference for quantile treatment effects based on analog estimators of our identifying formulas. 
While our identification result is novel, estimation and inference results follow from an adaptation of existing approaches to our framework.
Therefore, the main text focuses on the identification theory.
Details of estimation and inference theories are found in the appendix.

The rest of this paper is organized as follows. 
In Section \ref{sec:identification}, we develop the identification result. 
Section \ref{sec:methodology} presents a practical guideline on estimation and inference. 
Appendix \ref{sec:theory} presents formal theories for the method of inference. 
Appendix \ref{sec:additional} presents additional practical considerations.
Appendix \ref{sec:auxiliary_lemmas_and_proofs} contains mathematical details.

%%%%%%%%%%%%%%%%%%%%%%%%%%%%%%%%%%%%%%%%%%%%%%%%%%%%%%%%%%%%%%%%%%%%%%%%%%%%%%%%%%%%%%%%%%%%%%%
\section{Identification: the Main Result}
\label{sec:identification}

We model the random vector 
$(Y,D,X,U,V):(\Omega^{x},\mathscr{F}^{x},\mathds{P}^{x})\rightarrow\mathscr{Y}\times\mathscr{D}\times\mathscr{X}\times \mathscr{U}\times\mathscr{V}$ 
through the following causal structure, where 
$\mathscr{Y}\subset\mathbb{R}$, 
$\mathscr{D}=\{0,1\}$, 
$\mathscr{X}\subset\mathbb{R}$, 
$\mathscr{U}\subset\mathbb{R}^{d_{U}}$ for $d_{U}\in\mathbb{N}$,
and 
$\mathscr{V}\subset\mathbb{R}$.
\begin{align}
Y  &  =g(D,X,U)\label{eq:y}\\
D  &  =\mathbbm{1}\left\{  h(X)\geq V\right\}  \label{eq:d}
\end{align}
In equation (\ref{eq:y}), the outcome variable $Y$ is produced through function $g$ by a binary treatment variable $D$, a continuous running variable or assignment variable $X$, and miscellaneous factors $U$. 
We let $Y^{d}=g(d,X,U)$ denote the potential outcome random variable that an individual with attributes $(X,U)$ would produce under each hypothetical treatment choice $d\in\{0,1\}$. 
The actual treatment choice $D$ is determined by $X$ and $V$ through the threshold-crossing model (\ref{eq:d}). 
A researcher observes the joint distribution of $Y$, $D$, and $X$. 
However, a researcher cannot observe $U$ or $V$. 
We do \textit{not} impose any statistical independence condition in this model.
Therefore, existing methods for instrumental variable quantile regression \citep[e.g.,][]{chernozhukov_hansen2005} will not apply here. 
In particular, we do \textit{not} assume statistical independence between the running variable $X$ and the unobservables $(U,V)$. 
Instead, we make the following assumption of the regression kink design (RKD).

%%%%%%%%%%%%%%%%%%%%%%%%%%%%%%%%%%%%%%%%%%%%%%%%%%%%%%%%%%%%%%%%%%%%%%%%%%%%%%%%

\begin{assumption}
[Regression Kink Design, RKD]\label{a:rkd} 
Let $x_{0}=0\in\mathscr{X}$ be a designed kink location. 
\newline
(i) $h$ is continuously differentiable in a deleted neighborhood $I_{X}\backslash\{0\}\subset\mathscr{X}$ of $x_{0}=0$.
\newline
(ii) $h$ is continuous at $x_{0}=0$. \newline(iii) $\lim_{x\downarrow0}h^{\prime}(x)\neq\lim_{x\uparrow0}h^{\prime}(x)$, where
$h^{\prime}$ denotes $dh/dx$. 
\newline
(iv) The conditional distribution of $V$ given $X$ is absolutely continuous with a continuously differentiable conditional density function $f_{V|X}(\cdot|\cdot)$. 
\newline
(v) The conditional cumulative distribution function $F_{Y^{d}|VX}(y|\cdot,\cdot)$ is continuously differentiable for each $y\in\mathscr{Y}$ for each $d\in\{0,1\}$.
\newline
(vi) $f_{V|X}(h(0)|0)>0$.
\end{assumption}

%%%%%%%%%%%%%%%%%%%%%%%%%%%%%%%%%%%%%%%%%%%%%%%%%%%%%%%%%%%%%%%%%%%%%%%%%%%%%%%%

The research design as required by Assumption \ref{a:rkd} consists of three broad pieces. 
First, the treatment assignment rule $h$ has a kink at the designed location $x_{0}=0$, as formally stated in parts (ii) and (iii), but this assignment rule $h$ is reasonably smooth elsewhere, as formally stated in part (i). 
Second, every other function is reasonably smooth, as formally stated in parts (iv) and (v).
Third, there is sufficient data at the designed kink location $x_{0}=0$, as formally stated in part (vi). 
This assumption is analogous to that of \citet{dong2018} who analyzes average effects of binary treatments in the regression kink
design. 
Under this design, we obtain the following identification result for conditional distributions of the potential outcomes $Y^{d}$ given the event of $(V,X)=(h(0),0)$.

%%%%%%%%%%%%%%%%%%%%%%%%%%%%%%%%%%%%%%%%%%%%%%%%%%%%%%%%%%%%%%%%%%%%%%%%%%%%%%%%

\begin{theorem}
[Identification]\label{theorem:identification} Let Assumption \ref{a:rkd} hold for the model (\ref{eq:y})--(\ref{eq:d}). 
Then,
\begin{align*}
F_{Y^{1}|VX} (y |h(0), 0) =  &  \frac{\lim_{x \downarrow0} \frac{d}{dx}\E\left[  \mathbbm{1}\left\{  Y \leq y \right\}  \cdot D | X=x\right]  - \lim_{x \uparrow0} \frac{d}{dx}\E\left[  \mathbbm{1}\left\{  Y \leq y \right\}  \cdot D | X=x\right]  }{\lim_{x \downarrow0} \frac{d}{dx}\E\left[  D | X=x\right]  - \lim_{x \uparrow0} \frac{d}{dx}\E\left[  D | X=x\right]  }
\quad\text{and}\\
F_{Y^{0}|VX} (y |h(0), 0) =  &  \frac{\lim_{x \downarrow0} \frac{d}{dx}\E\left[  \mathbbm{1}\left\{  Y \leq y \right\}  \cdot(1-D) | X=x\right] - \lim_{x \uparrow0} \frac{d}{dx}\E\left[  \mathbbm{1}\left\{  Y \leq y \right\}  \cdot(1-D) | X=x\right]  }{\lim_{x \downarrow0} \frac{d}{dx}\E\left[  1-D | X=x\right]  - \lim_{x \uparrow0} \frac{d}{dx}\E\left[  1-D | X=x\right]  }
\end{align*}
hold for all $y \in\mathscr{Y}$.
\end{theorem}

%%%%%%%%%%%%%%%%%%%%%%%%%%%%%%%%%%%%%%%%%%%%%%%%%%%%%%%%%%%%%%%%%%%%%%%%%%%%%%%%

Once the conditional cumulative distribution functions, $F_{Y^{d}|VX}(\cdot|h(0),0)$ for $d\in\{0,1\}$, are identified through the formulas presented in Theorem \ref{theorem:identification}, the conditional quantile treatment effect is in turn identified by
\begin{equation}
\tau(\theta)=\inf\{y\in\mathscr{Y}:F_{Y^{1}|VX}(y|h(0),0)\geq\theta\}-\inf\{y\in\mathscr{Y}:F_{Y^{0}|VX}(y|h(0),0)\geq\theta\} \label{eq:qte}
\end{equation}
for $\theta\in(0,1)$. 
Theorem \ref{theorem:identification}
also provides the identification of the distributional treatment effects for local complies, $F_{Y^{1}|VX}(\cdot|h(0),0)-F_{Y^{0}|VX}(\cdot|h(0),0)$, as in \citet{abadie2002} and \citet{shen_zhang2016}, which are useful to test important hypotheses such as the first order stochastic dominance.\footnote{We remark that, with our identifying formulas provided in Theorem \ref{theorem:identification}, $F_{Y^{1}|VX}(\cdot|h(0),0)-F_{Y^{0}|VX}(\cdot|h(0),0)$ can be simply expressed as a single Wald ratio: $\frac{\lim_{x \downarrow0} \frac{d}{dx}\E\left[  \mathbbm{1}\left\{  Y \leq y \right\} | X=x\right]  - \lim_{x \uparrow0} \frac{d}{dx}\E\left[  \mathbbm{1}\left\{  Y \leq y \right\}  | X=x\right]  }{\lim_{x \downarrow0} \frac{d}{dx}\E\left[  D | X=x\right]  - \lim_{x \uparrow0} \frac{d}{dx}\E\left[  D | X=x\right]  }$.}\bigskip\newline
%%%%%%%%%%%%%%%%%%%%%%%%%%%%%%%%%%%%%%%%%%%%%%%%%%%%%%%%%%%%%%%%%%%%%%%%%%%%%%%%
\textbf{Proof of Theorem \ref{theorem:identification}:} 
By applying Leibniz rule under Assumption
\ref{a:rkd} (i) and (iv), we have
\[
\frac{d}{dx}\E\left[  D|X=x\right]  =\frac{d}{dx}\int_{-\infty}^{h(x)} f_{V|X}(v|x)dv=h^{\prime}(x)\cdot f_{V|X}(h(x)|x)+\int_{-\infty}^{h(x)}\frac{\partial}{\partial x}f_{V|X}(v|x)dv
\]
for all $x\in I_{X}\backslash\{0\}$. 
Similarly, by applying Leibniz rule under Assumption \ref{a:rkd} (i), (iv), and (v), we have
\begin{align*}
\frac{d}{dx}\E\left[  \mathbbm{1}\left\{  Y\leq y\right\}  \cdot D|X=x\right] 
=  &  \frac{d}{dx}\int_{-\infty}^{h(x)}\int_{u:g(1,x,u)\leq y}F_{UV|X} (du,dv|x)
\\
=  &  \frac{d}{dx}\int_{-\infty}^{h(x)}f_{V|X}(v|x)\int_{u:g(1,x,u)\leq y}F_{U|VX}(du|v,x)dv
\\
=  &  \frac{d}{dx}\int_{-\infty}^{h(x)}f_{V|X}(v|x)\cdot F_{Y^{1} |VX}(y|v,x)dv
\\
=  &  h^{\prime}(x)\cdot f_{V|X}(h(x)|x)\cdot F_{Y^{1}|VX}(y|h(x),x)+
\\
&  \int_{-\infty}^{h(x)}\frac{d}{dx}\left[  f_{V|X}(v|x)\cdot F_{Y^{1} |VX}(y|v,x)dv\right]  dv
\end{align*}
for all $(x,y)\in\left(  I_{X}\backslash\{0\}\right)  \times\mathscr{Y}$.
Therefore, by Assumption \ref{a:rkd} (ii) and (iv), we can write
\[
\lim_{x\downarrow0}\frac{d}{dx}\E\left[  D|X=x\right]  -\lim_{x\uparrow0}\frac{d}{dx}\E\left[  D|X=x\right]  
=
\left[  h^{\prime}(0^{+})-h^{\prime }(0^{-})\right]  \cdot f_{V|X}(h(0)|0),
\]
and, by Assumption \ref{a:rkd} (ii), (iv), and (v), we can write
\begin{align*}
&  \lim_{x\downarrow0}\frac{d}{dx}\E\left[  \mathbbm{1}\left\{  Y\leq y\right\}  \cdot D|X=x\right]  -\lim_{x\uparrow0}\frac{d}{dx}\E\left[ \mathbbm{1}\left\{  Y\leq y\right\}  \cdot D|X=x\right] 
\\
=  
&  \left[  h^{\prime}(0^{+})-h^{\prime}(0^{-})\right]  \cdot f_{V|X} (h(0)|0)\cdot F_{Y^{1}|VX}(y|h(0),0)
\end{align*}
for all $y\in\mathscr{Y}$. 
Taking the ratio of these expressions under Assumption \ref{a:rkd} (iii) and (vi) yields
\[
\frac{\lim_{x\downarrow0}\frac{d}{dx}\E\left[  \mathbbm{1}\left\{  Y\leq y\right\}  \cdot D|X=x\right]  -\lim_{x\uparrow0}\frac{d}{dx}\E\left[ \mathbbm{1}\left\{  Y\leq y\right\}  \cdot D|X=x\right]  }{\lim_{x\downarrow 0}\frac{d}{dx}\E\left[  D|X=x\right]  -\lim_{x\uparrow0}\frac{d}{dx}\E\left[ D|X=x\right]  }=F_{Y^{1}|VX}(y|h(0),0)
\]
for all $y\in\mathscr{Y}$. Similar lines of arguments yield
\begin{align*}
\frac{\lim_{x\downarrow0}\frac{d}{dx}\E\left[  \mathbbm{1}\left\{  Y\leq y\right\}  \cdot(1-D)|X=x\right]  -\lim_{x\uparrow0}\frac{d}{dx}\E\left[ \mathbbm{1}\left\{  Y\leq y\right\}  \cdot(1-D)|X=x\right]  }{\lim_{x\downarrow0}\frac{d}{dx}\E\left[  1-D|X=x\right]  -\lim_{x\uparrow0} \frac{d}{dx}\E\left[  1-D|X=x\right]  }
\\
=F_{Y^{0}|VX}(y|h(0),0)
\end{align*}
for all $y\in\mathscr{Y}$. 
\qed
%%%%%%%%%%%%%%%%%%%%%%%%%%%%%%%%%%%%%%%%%%%%%%%%%%%%%%%%%%%%%%%%%%%%%%%%%%%%%%%%
\bigskip\newline\textbf{Discussions of Theorem \ref{theorem:identification}:}
In the context of the regression discontinuity design (RDD) where $h(0^{-})<h(0^{+})$, \citet{frandsen_frolich_melly2012} show that similar local Wald ratios identify the conditional distribution of the potential outcomes given the event 
$$
C_{\text{RDD}}=\{\omega\in\Omega:X(\omega )=0,h(0^{-})<V(\omega)\leq h(0^{+})\}
$$
of local compliance. 
In our context of the regression kink design where $h(0^{-})=h(0^{+})$, Theorem \ref{theorem:identification} shows that local Wald ratios of the derivatives identify the conditional distributions of the potential outcomes given the event 
$$
C_{\text{RKD}}=\{\omega\in\Omega:X(\omega)=0,V(\omega)=h(0)\},
$$
which may be considered as a limit of the event $C_{\text{RDD}}$ for RDD as
$\left\vert h(0^{+})-h(0^{-})\right\vert $ approaches $0$. 
In this sense, our causal interpretation result is similar to that of the marginal treatment effects \citep{bjorklund_moffitt1987,heckman_vytlacil1999,heckman_vytlacil2005}.
This interpretation is analogous to the identification result by \citet{dong2018} who analyzes average effects of binary treatments in the regression kink design.
$\triangle$
%%%%%%%%%%%%%%%%%%%%%%%%%%%%%%%%%%%%%%%%%%%%%%%%%%%%%%%%%%%%%%%%%%%%%%%%%%%%%%%%%%%%%%%%%%%%%%%
\section{Estimation and Inference: a Practical Guideline}
\label{sec:methodology}

While the main contribution of this paper lies in our new identification result presented in Section \ref{sec:identification}, we also develop a theory and method of estimation and inference for completeness.
Since the estimation and inference strategies are standard, we relegate most of the details to the appendix.
In this section, we present a practical guideline on estimation and inference for the conditional quantile treatment effects $\tau(\theta)$. 
A formal theory is presented in Appendix \ref{sec:theory}.
We also present additional practical considerations in Appendix \ref{sec:additional}.
Auxiliary lemmas and proofs are found Appendix \ref{sec:auxiliary_lemmas_and_proofs}.

The local Wald ratios proposed in Theorem \ref{theorem:identification} as identifying formulae can be succinctly rewritten as
\begin{equation}
{F}_{Y^{d}|VX}(y|h(0),0)=\frac{{\mu}_{1}^{\prime}(0^{+},y,d)-{\mu}_{1}^{\prime}(0^{-},y,d)}{{\mu}_{2}^{\prime}(0^{+},d)-{\mu}_{2}^{\prime}(0^{-},d)}, \label{eq:local_wald_estimand}
\end{equation}
where $\mu_{1}^{\prime}(x,y,d)$ and $\mu_{2}^{\prime}(x,d)$ are the partial derivatives with respect to $x$ of $\mu_{1}(x,y,d)$ and $\mu_{2}(x,d)$ defined by
\begin{align*}
\mu_{1}(x,y,d)  &  =E[\mathds{1}\{Y\leq y\}\cdot\mathds{1}\{D=d\}|X=x]
\qquad \text{and}\\
\mu_{2}(x,d)  &  =E[\mathds{1}\{D=d\}|X=x], 
\end{align*}
respectively. We estimate the components of (\ref{eq:local_wald_estimand}) by the one-sided local cubic estimators
\begin{align}
&  \hat{\mu}_{1}^{\prime}(0^{\pm},y,d)h_{n}=e_{1}^{\top}\mathop{\rm arg~min}\limits_{\alpha\in\mathds{R}^{4}}\sum_{i=1}^{n}\Big[\mathds{1}\{Y_{i}\leq y\}\mathds{1}\{D_{i}=d\}-r_{3}^{\top}\Big(\frac{X_{i}}{h_{n}}\Big)\alpha\Big]^{2}K\Big(\frac{X_{i}}{h_{n}}\Big)\delta_{i}^{\pm}
\quad\text{and}\label{eq:local_slope_estimator1}\\
&  \hat{\mu}_{2}^{\prime}(0^{\pm},d)h_{n}=e_{1}^{\top}\mathop{\rm arg~min}\limits_{\alpha\in\mathds{R}^{4}}\sum_{i=1}^{n}\Big[\mathds{1}\{D_{i}=d\}-r_{3}^{\top}\Big(\frac{X_{i}}{h_{n}}\Big)\alpha\Big]^{2}K\Big(\frac{X_{i}}{h_{n}}\Big)\delta_{i}^{\pm},
\label{eq:local_slope_estimator2}
\end{align}
where $K$ is a kernel function, $h_{n}$ is a bandwidth parameter, $e_{1}=(0,1,0,0)^{\top}$, $r_{3}(u)=(1,u,u^{2},u^{3})^{\top}$, $\delta_{i}^{+}=\mathds{1}\{X_{i}\geq0\}$ and $\delta_{i}^{-}=\mathds{1}\{X_{i}<0\}$.
A plug-in estimator for (\ref{eq:local_wald_estimand}) is given by
\[
\widehat{F}_{Y^{d}|VX}(y|h(0),0)=\frac{\hat{\mu}_{1}^{\prime}(0^{+},y,d)-\hat{\mu}_{1}^{\prime}(0^{-},y,d)}{\hat{\mu}_{2}^{\prime}(0^{+},d)-\hat{\mu}_{2}^{\prime}(0^{-},d)}.
\]
The motivation for our using the local cubic polynomial is to account for the manual bias correction from local quadratic estimators. 
By considering the asymptotic distribution for the higher-order local polynomial, we effectively account for bias estimation in the asymptotic distribution from the lower-order one, thus allowing for robustness in inference against large bandwidths -- see Calonico, Cattaneo and Titiunik (2014, Remark 7) and Remark S.A.7 in their supplementary material.

We can in turn estimate the conditional quantile treatment effect $\tau(\theta)$ by
\begin{align*}
\hat{\tau}(\theta)=  &  \inf\left\{  y\in\mathscr{Y}:\widehat{F}_{Y^{1}|VX}(y|h(0),0)\geq\theta\right\}  -\inf\left\{  y\in\mathscr{Y}:\widehat{F}_{Y^{0}|VX}(y|h(0),0)\geq\theta\right\} 
\\
&  =\widehat{Q}_{Y^{1}|VX}(\theta)-\widehat{Q}_{Y^{0}|VX}(\theta).
\end{align*}
The local Wald estimator $\widehat{F}_{Y^{d}|VX}(\cdot|h(0),0)$ is not always monotone increasing in finite sample. 
For ease of implementing the CDF inversion, we monotonize the estimated CDFs by re-arrangements following \citet{chernozhukov_fernandez-val_galichon2016}. 
This does not affect the asymptotic properties of the estimators, while allowing for inversion of the CDF estimators. \citet{frandsen_frolich_melly2012} also use this
technology in the context of the regression discontinuity design.

Let $\Gamma^{\pm}=\int_{\mathds{R}_{\pm}}r_{3}(u)r_{3}^{\top}(u)K(u)du$. 
Under the assumptions to be stated in Appendix \ref{sec:theory}, we obtain the following Uniform Bahadur Representations (BR) for the local slope estimators (\ref{eq:local_slope_estimator1}) and (\ref{eq:local_slope_estimator2}).
\begin{align}
\nu_{n}^{\pm}(y,d,1)=  &  \sqrt{nh_{n}^{3}}[\hat{\mu}_{1}^{\prime}(0^{\pm},y,d)-\mu_{1}^{\prime}(0^{\pm},y,d)+O_{p}\left(  h_{n}^{3}\right) ]
\label{eq:br1}
\\
=  &  \frac{1}{\sqrt{nh_{n}}f_{X}(0)}\sum_{i=1}^{n}e_{1}^{\top}(\Gamma^{\pm })^{-1}r_{3}\Big(\frac{X_{i}}{h_{n}}\Big)\Big[\mathds{1}\{Y_{i}\leq y\}\mathds{1}\{D_{i}=d\}-\mu_{1}(X_{i},y,d)\Big]K\Big(\frac{X_{i}}{h_{n} }\Big)\delta_{i}^{\pm}
\nonumber
\\
\nu_{n}^{\pm}(y,d,2)=  &  \sqrt{nh_{n}^{3}}[\hat{\mu}_{2}^{\prime}(0^{\pm },d)-\mu_{2}^{\prime}(0^{\pm},d)+O_{p}\left(  h_{n}^{3}\right) ]
\label{eq:br2}
\\
=  &  \frac{1}{\sqrt{nh_{n}}f_{X}(0)}\sum_{i=1}^{n}e_{1}^{\top}(\Gamma^{\pm })^{-1}r_{3}\Big(\frac{X_{i}}{h_{n}}\Big)\Big[\mathds{1}\{D_{i}=d\}-\mu _{2}(X_{i},d)\Big]K\Big(\frac{X_{i}}{h_{n}}\Big)\delta_{i}^{\pm}
\nonumber
\end{align}
We note that $\nu_{n}^{\pm}(y,d,2)$ are trivial functions of $y$.

Covariance functions for the limit processes are often cumbersome to approximate in practice.
\citet{qu_yoon2018} propose a simulation method to approximate limit processes under sharp designs -- also see \citet{qu_yoon2015} -- but this method is not applicable to fuzzy designs.
We thus propose to use the multiplier bootstrap method to approximate the asymptotic distributions of these BR. 
Draw a random sample $\xi_{1},...,\xi_{n}$ from the standard normal distribution independently from the data $\{Y_{i},D_{i},X_{i}\}_{i=1}^{n}$. 
Replacing the unknowns $\mu_{1},\mu_{2}$ and $f_{X}(0)$ in the BR by their uniformly consistent estimators $\tilde{\mu}_{1},\tilde
{\mu}_{2}$ and $\hat{f}_{X}\left(  0\right)  $, respectively, we define the following Estimated Multiplier Processes (EMP).
\begin{align}
\hat{\nu}_{\xi,n}^{\pm}(y,d,1)=  &  \frac{1}{\sqrt{nh_{n}}\hat{f}_{X}(0)} \sum_{i=1}^{n}\xi_{i}e_{1}^{\top}(\Gamma^{\pm})^{-1}r_{3}\Big(\frac{X_{i} }{h_{n}}\Big)\Big[\mathds{1}\{Y_{i}\leq y\}\mathds{1}\{D_{i}=d\}-\tilde{\mu }_{1}(X_{i},y,d)\Big]K\Big(\frac{X_{i}}{h_{n}}\Big)\delta_{i}^{\pm }
\label{eq:esp1}
\\
\hat{\nu}_{\xi,n}^{\pm}(y,d,2)=  &  \frac{1}{\sqrt{nh_{n}}\hat{f}_{X}(0)} \sum_{i=1}^{n}\xi_{i}e_{1}^{\top}(\Gamma^{\pm})^{-1}r_{3}\Big(\frac{X_{i} }{h_{n}}\Big)\Big[\mathds{1}\{D_{i}=d\}-\tilde{\mu}_{2}(X_{i} ,d)\Big]K\Big(\frac{X_{i}}{h_{n}}\Big)\delta_{i}^{\pm}
\label{eq:esp2}
\end{align}

Under the assumptions to be stated in Appendix \ref{sec:theory}, we show that the EMP can be used to uniformly approximate the asymptotic distribution of the BR. 
Consequently, by the functional delta method, the asymptotic distribution of
\[
\sqrt{nh_{n}^{3}}[\hat{\tau}(\cdot)-\tau(\cdot)]
\]
can be approximated uniformly on $\Theta=[a,1-a]$ for $a\in(0,1/2)$ by the estimated process
\[
\widehat{\Xi}(\cdot)=-\left[  \frac{\hat{Z}_{\xi,n}(\hat{Q}_{Y^{1}|VX} (\cdot),1)}{\hat{f}_{Y^{1}|VX}(\hat{Q}_{Y^{1}|VX}(\cdot)|h(0),0)}-\frac {\hat{Z}_{\xi,n}(\hat{Q}_{Y^{0}|VX}(\cdot),0)}{\hat{f}_{Y^{0}|VX}(\hat {Q}_{Y^{0}|VX}(\cdot)|h(0),0)}\right]  ,
\]
where
\begin{align*}
&  \hat{Z}_{\xi,n}(y,d)=\\
&  \frac{[\hat{\mu}_{2}^{\prime}(0^{+},d)-\hat{\mu}_{2}^{\prime} (0^{-},d)][\hat{\nu}_{\xi,n}^{+}(y,d,1)-\hat{\nu}_{\xi,n}^{-}(y,d,1)]-[\hat {\mu}_{1}^{\prime}(0^{+},y,d)-\hat{\mu}_{1}^{\prime}(0^{-},y,d)][\hat{\nu }_{\xi,n}^{+}(y,d,2)-\hat{\nu}_{\xi,n}^{-}(y,d,2)]}{[\hat{\mu}_{2}^{\prime }(0^{+},d)-\hat{\mu}_{2}^{\prime}(0^{-},d)]^{2}}.
\end{align*}

Once we obtain these approximations to the asymptotic distributions, we may conduct various tests of quantile functions following \citet{koenker_xiao2002} and \citet{chernozhukov_fernandez-val2005}. 
For example for the test of treatment significance, we use the test statistic
\[
T^{TS} = \sup_{\theta\in\Theta} \left\vert \sqrt{nh_{n}^{3}} \hat\tau
(\theta)\right\vert
\]
where $\Theta= [a,1-a]$ for some $a \in(0,1/2)$. 
We can approximate the asymptotic distribution of $T^{TS}$ by
\[
\sup_{\theta\in\Theta} \left\vert \widehat{\Xi}(\theta)\right\vert .
\]
Similarly, for the test of treatment homogeneity, we use the test statistic
\[
T^{TH} = \sup_{\theta\in\Theta} \left\vert \sqrt{nh_{n}^{3}} \left(  \hat \tau(\theta) - \int_{\Theta}\hat\tau(\vartheta) d\vartheta\right)  \right\vert.
\]
We can approximate the asymptotic distribution of $T^{TH}$ by
\[
\sup_{\theta\in\Theta} \left\vert \left(  \widehat{\Xi}(\theta) - \int%
_{\Theta}\widehat{\Xi}(\vartheta) d\vartheta\right)  \right\vert .
\]

In this section, we presented a practical guideline on estimation and inference for the conditional quantile treatment effects $\tau(\theta)$. 
We refer interested readers to Appendix \ref{sec:theory} for a formal theory.
Furthermore, Appendix \ref{sec:additional} presents additional practical considerations not covered in this section.

\section{Summary}

\label{sec:summary}

The existing literature on identification in regression kink designs includes the following three results.
\citet{card_lee_pei_weber2015} propose identification of average effects of continuous treatments.
\citet{dong2018} proposes identification of average effects of binary treatments.
\citet{chiang_sasaki2019} propose identification of quantile-wise effects of continuous treatments. 
On the other hand, this literature has been missing an identification result for quantile-wise effects of binary treatments.
To complete this literature on identification, we propose identification of quantile-wise effects of binary treatments in this paper in regression kink designs.

Specifically, we show that a local Wald ratio of derivatives of certain conditional expectation functions identifies the conditional distribution functions of potential outcomes given the event of local compliance.
Taking the difference of the left-inverses of these identified conditional distribution functions in turn identifies the conditional quantile treatment effects given the event of local compliance. 
While the main contribution of this paper is the identification result, we also develop a theory and method of estimation and inference for completeness.

%%%%%%%%%%%%%%%%%%%%%%%%%%%%%%%%%%%%%%%%%%%%%%%%%%%%%%%%%%%%%%%%%%%%%%%%%%%%%%%%%%%%%%%%%%%%%%%
\appendix
%%%%%%%%%%%%%%%%%%%%%%%%%%%%%%%%%%%%%%%%%%%%%%%%%%%%%%%%%%%%%%%%%%%%%%%%%%%%%%%%%%%%%%%%%%%%%%%

\newpage
\section*{Mathematical Appendix}

%%%%%%%%%%%%%%%%%%%%%%%%%%%%%%%%%%%%%%%%%%%%%%%%%%%%%%%%%%%%%%%%%%%%%%%%%%%%%%%%%%%%%%%%%%%%%%%

\section{Estimation and Inference: Formal Theory}

\label{sec:theory}
%%%%%%%%%%%%%%%%%%%%%%%%%%%%%%%%%%%%%%%%%%%%%%%%%%%%%%%%%%%%%%%%%%%%%%%%%%%%%%%%%%%%%%%%%%%%%%%

We use the following set of assumptions for the uniform Bahadur Representations, the bootstrap
validity, and consistent conditional density and first-stage estimations. 
Fix
$a\in(0,1/2)$ and 
$\epsilon>0$, denote 
$$
\mathscr{Y}_{1}=[Q_{Y^{1}%
|VX}(a)-\epsilon,Q_{Y^{1}|VX}(1-a)+\epsilon]\cup\lbrack Q_{Y^{0}%
|VX}(a)-\epsilon,Q_{Y^{0}|VX}(1-a)+\epsilon].
$$ 
We will write $a\lesssim b$ if
there exists a universal constant $C$ such that $a\leq Cb$. 
Denote
$$
P_{D|X}(d|x)=\mathds{P}^{x}(D=d|X=x).
$$
We define the following objects for all $y_{1}$, $y_{2}%
\in\mathscr{Y}_{1}$, $d_{1}$, $d_{2}\in\mathscr{D}:$
\begin{align*}
\sigma_{11}((y_{1},d_{1}),(y_{2},d_{2})|x)  &  =E[(\mathds{1}\{Y\leq y_{1},D=d_{1}\}-\mu_{1}(X,y_{1},d_{1})) \cdot 
\\
& \qquad \: \: (\mathds{1}\{Y\leq y_{2},D=d_{2}\}-\mu_{1}(X,y_{2},d_{2}))|X=x],\\
\sigma_{22}((y_{1},d_{1}),(y_{2},d_{2})|x)  &  =E[(\mathds{1}\{D=d_{1}%
\}-\mu_{2}(X,d_{1}))(\mathds{1}\{D=d_{2}\}-\mu_{2}(X,d_{2}))|X=x],\qquad\text{and}\\
\sigma_{12}((y_{1},d_{1}),(y_{2},d_{2})|x)  &  =E[(\mathds{1}\{Y\leq
y_{1},D=d_{1}\}-\mu_{1}(X,y_{1},d_{1}))(\mathds{1}\{D=d_{2}\}-\mu_{2}%
(X,d_{2}))|X=x].
\end{align*}
%%%%%%%%%%%%%%%%%%%%%%%%%%%%%%%%%%%%%%%%%%%%%%%%%%%%%%%%%%%%%%%%%%%%%%%%%%%%%%%%%%%%%%%%%%%%%%%

\begin{assumption}
\label{a:inference} Let $[\underline{x},\overline{x}]$ be a compact interval
containing $0$ in its interior. Let $a\in(0,1/2)$.\newline(i) (a)
$\{Y_{i},D_{i},X_{i}\}_{i=1}^{n}$ are $n$ independent copies of random vector
$(Y,D,X)$ with support $\mathscr{Y}\times\mathscr{D}\times\mathscr{X}$ defined
on a probability space $(\Omega^{x},\mathscr{F}^{x},\mathds{P}^{x})$. (b) $X$
has a continuously differentiable density function $f_{X}$ with $0<f_{X}%
(0)<\infty$.
%(c) $F_{Y|DX}(\cdot|d,\cdot)$ has continuous second partial derivatives on $\mathscr{Y}_1\times [\underline{x},\overline{x}]$.
(c) $f_{YD|X}(y,d|x)$ is well-defined on $\mathscr{Y}_{1}\times
\mathscr{D}\times([\underline{x},\overline{x}]\setminus\{0\})$ and
$|f_{YD|X}(y,d|0^{+})-f_{YD|X}(y,d|0^{-})|>m>0$ on $\mathscr{Y}_{1}%
\times\mathscr{D}$. \newline(ii)(a) Conditional density $f_{Y|XD}$ is
Lipschitz continuous on $\mathscr{Y}_{1}\times\lbrack\underline{x},0)$ and
$\mathscr{Y}_{1}\times(0,\overline{x}]$ for each $d$ and is four-time
partially differentiable in $x$ and twice partially differentiable in $y$ for
each $d$. $\frac{\partial^{j}}{\partial x^{j}}\frac{\partial^{k}}{\partial
y^{k}}f_{Y|XD}(\cdot|\cdot,d)$ is continuous and uniformly bounded on
$\mathscr{Y}_{1}\times\lbrack\underline{x},0)$ and $\mathscr{Y}_{1}%
\times(0,\overline{x}]$ for each $d$ for all $j$, $k\in\mathds{N}$, $j+k\leq
4$. (b) $P_{D|X}(d|\cdot)$ is Lipschitz continuous in $x$, four-time
differentiable on $[\underline{x},0)$ and $(0,\overline{x}]$ for each $d$.
$\frac{\partial^{4}}{\partial x^{4}}P_{D|X}(d|\cdot)$ is continuous and
uniformly bounded on $[\underline{x},0)$ and $(0,\overline{x}]$ for each d.
(c) For any $y_{1}$, $y_{2}\in\mathscr{Y}_{1}$, $d_{1}$, $d_{2}\in
\mathscr{D}$, we have $\sigma_{11}((y_{1},d_{1}),(y_{2},d_{2})|\cdot)$,
$\sigma_{12}((y_{1},d_{1}),(y_{2},d_{2})|\cdot)$ and $\sigma_{22}((y_{1}%
,d_{1}),(y_{2},d_{2})|\cdot)\in\mathcal{C}^{1}([\underline{x},\overline
{x}]\setminus\{0\})$ where $\mathcal{C}^{1}$ is the collection of continuously
differentiable functions.\newline(iii) The bandwidths satisfy $h_{n}%
\rightarrow0$, $nh_{n}^{3}\rightarrow\infty$, $nh_{n}^{9}\rightarrow0$,
$0<h_{n}\leq h_{0}$ for some finite $h_{0}$.\newline(iv) (a)
$K:[-1,1]\rightarrow\mathds{R}_{+}$ is bounded and $\int_{\mathds{R}}%
K(u)du=1$. (b) $\{K(\cdot/h):h>0\}$ is of VC type. (c) $\Gamma^{\pm}%
=\int_{\mathds{R}_{\pm}}r_{3}(u)r_{3}^{\top}(u)K(u)du$ are positive
definite.\newline(v) $\hat{f}_{X}(0)$ is a consistent estimator for $f_{X}%
(0)$. For $d=0,1$, $\hat{f}_{Y^{d}|VX}(\cdot|h(0),0)$ are uniformly consistent
estimators for $f_{Y^{d}|VX}(\cdot|h(0),0)$. $\tilde{\mu}_{1}%
(x,y,d)\mathds{1}\{|x/h_{n}|\leq1\}$ and $\tilde{\mu}_{2}%
(x,d)\mathds{1}\{|x/h_{n}|\leq1\}$ are uniformly consistent estimators for
$\mu_{1}(x,y,d)\mathds{1}\{|x/h_{n}|\leq1\}$ and $\mu_{2}%
(x,d)\mathds{1}\{|x/h_{n}|\leq1\}$ on $\mathscr{X}\times\mathscr{Y}_{1}%
\times\mathscr{D}$. \newline(vi) $\{\xi_{1},...,\xi_{n}\}$ are $n$ independent
and identically distributed copies of a standard normal random variable $\xi$
defined on a probability space $(\Omega^{\xi},\mathscr{F}^{\xi}%
,\mathds{P}^{\xi})$ that is independent of $(\Omega^{x},\mathscr{F}^{x}%
,\mathds{P}^{x})$.
\end{assumption}

%%%%%%%%%%%%%%%%%%%%%%%%%%%%%%%%%%%%%%%%%%%%%%%%%%%%%%%%%%%%%%%%%%%%%%%%%

Part (i) concerns about the sampling procedure and the distribution of data.
Part (ii) requires smoothness of the conditional expectation functions on a
deleted neighborhood of $x_{0}=0$. Part (iii) regulates the rate at which
bandwidth decreases, which is consistent with examples of common choice rules
to be presented in Appendix \ref{subsec:bandwidth}. 
For example, the MSE-optimal bandwidth for the local quadratic estimator (e.g., $nh_{n}^{7}\rightarrow\infty$) is allowed. 
%Nevertheless, $nh_{n}%
%^{9}\rightarrow0$\textbf{ rules out the MSE-optimal bandwidth under the
%local cubic estimators, but it is fine due to the fact we treat local slope
%estimates from the (not bias-corrected) local cubic regression as the
%bias-corrected counterparts from the local quadratic one. }
Part (iv) is
satisfied by common kernel functions, such as uniform, triangular, biweight,
triweight, and Epanechnikov kernels, for example. Part (v) is a high-level
assumption of (uniformly) consistent estimation of the first-stage estimators.
While we keep this high-level statement for the current section, Appendix
\ref{subsec:first_stage} proposes concrete examples of such uniformly
consistent estimators. Part (vi) requires the multiplier random sample to be
drawn independently of the data $\{Y_{i},D_{i},X_{i}\}_{i=1}^{n}$. We remark
that part (vi) implies that all (uniformly) consistent estimators with respect
to $\mathds{P}^{x}$ are also (uniformly) consistent with respect to
$\mathds{P}^{x\times\xi}$.

Under Assumption \ref{a:inference} (i), (ii)(a)(b), (iii), (iv), an
application of Lemma 1 of \citet{chiang_hsu_sasaki2019} gives the uniform
Bahadur Representation as in equations (\ref{eq:br1}) and (\ref{eq:br2}). The
following theorem establishes (i) (a) the asymptotic distribution of the BR;
(i) (b) the asymptotic distribution of the local Wald estimators; (i) (c) the
asymptotic distribution of the conditional quantile treatment effect
estimator; and (ii) the bootstrap validity. A proof is provided in Appendix
\ref{sec:MB}.

\begin{theorem}
[Asymptotic Distributions and Bootstrap Validity]\label{theorem:MB} Suppose
Assumptions \ref{a:rkd} and \ref{a:inference} hold, then there exists a zero
mean Gaussian process $\mathds{G}:\Omega^{x}\mapsto\ell^{\infty}%
(\{\mathscr{Y}_{1}\times\mathscr{D}\times\{1,2\}\})$, where $l^{\infty}$ is
the collection of all bounded real valued functions, such that:\newline(i) (a)
$\nu_{n}^{+}-\nu_{n}^{-}\leadsto\mathds{G}$. \newline(i) (b) $\sqrt{nh_{n}%
^{3}}[\widehat{F}_{Y^{d}|VX}(\cdot|h(0),0)-F_{Y^{d}|VX}(\cdot|h(0),0)]\leadsto
\mathds{G}_{F}(\cdot,d)$ holds, where $\mathds{G}_{F}(\cdot,d)$ is given by
\[
\mathds{G}_{F}(y,d)=\frac{[\mu_{2}^{\prime}(0^{+},d)-\mu_{2}^{\prime}%
(0^{-},d)]\mathds{G}(y,d,1)-[\mu_{1}^{\prime}(0^{+},y,d)-\mu_{1}^{\prime
}(0^{-},y,d)]\mathds{G}(y,d,2)}{[\mu_{2}^{\prime}(0^{+},d)-\mu_{2}^{\prime
}(0^{-},d)]^{2}}.
\]
(i) (c) $\sqrt{nh_{n}^{3}}[\hat{\tau}-\tau]\leadsto\mathds{G}_{\tau}$ holds,
where $\mathds{G}_{\tau}$ is given for each $\theta\in\Theta=[a,1-a]$ by
\[
\mathds{G}_{\tau}(\theta)=-\left[  \frac{\mathds{G}_{F}(Q_{Y^{1}|VX}%
(\theta|h(0),0),1)}{f_{Y^{1}|VX}(Q_{Y^{1}|VX}(\theta|h(0),0)|h(0),0)}%
-\frac{\mathds{G}_{F}(Q_{Y^{0}|VX}(\theta|h(0),0),1)}{f_{Y^{0}|VX}%
(Q_{Y^{0}|VX}(\theta|h(0),0)|h(0),0)}\right]  .
\]
(ii) We have
\[
\widehat{\Xi}(\cdot)=-\left[  \frac{\hat{Z}_{\xi,n}(\hat{Q}_{Y^{1}|VX}%
(\cdot|h(0),0),1)}{\hat{f}_{Y^{1}|VX}(\hat{Q}_{Y^{1}|VX}(\cdot
|h(0),0)|h(0),0)}-\frac{\hat{Z}_{\xi,n}(\hat{Q}_{Y^{0}|VX}(\cdot
|h(0),0),0)}{\hat{f}_{Y^{0}|VX}(\hat{Q}_{Y^{0}|VX}(\cdot|h(0),0)|h(0),0)}%
\right]  \underset{\xi}{\overset{p}{\leadsto}}\mathds{G}_{\tau}(\cdot)
\]

\end{theorem}

\begin{remark}
By considering the asymptotic distribution for the local cubic local
polynomial above, we effectively account for bias estimation in the asymptotic
distribution from the local quadratic kernel estimate-- see Calonico, Cattaneo
and Titiunik (2014, Remark 7) and Remark S.A.7 in their supplementary
material. Therefore, the proposed theory and bootstrap allow for robust
inference under the MSE-optimal bandwidth from the local quadratic kernel estimate.
\end{remark}

\begin{remark}\label{remark:equivalence}
$\hat{\mu}_{1}^{\prime}(0^{\pm},y,d)$, $\hat{\mu}_{2}^{\prime}(0^{\pm},d)$ and
Theorem 2 are developed for the unconstrained estimators, that is, without
imposing continuity in the conditional expectation of $\mathds{1}\{Y_{i}\leq
y\}\mathds{1}\{D_{i}=d\}$ and $\mathds{1}\{D_{i}=d\}$. On the other hand, for
example, consider the constrained version with the restriction with $\mu
_{1}(0^{+},y,d)=\mu_{1}(0^{-},y,d)$: the estimates can be obtained by solving
the \textquotedblleft pooled\textquotedblright\ least squares problem
\[
\mathop{\rm arg~min}\limits_{\left\{  \alpha,b^{+},b^{-}\right\}
\in\mathds{R}^{7}}\sum_{i=1}^{n}\Big[\mathds{1}\{Y_{i}\leq
y\}\mathds{1}\{D_{i}=d\}-\alpha-\delta_{i}^{+}r_{3\backslash0}^{\top
}\Big(\frac{X_{i}}{h_{n}}\Big)b^{+}-\delta_{i}^{-}r_{3\backslash0}^{\top
}\Big(\frac{X_{i}}{h_{n}}\Big)b^{-}\Big]^{2}K\Big(\frac{X_{i}}{h_{n}}\Big)
\]
where $r_{3\backslash0}(u)=\left(  u,u^{2},u^{3}\right)  $ and $b^{\pm}%
\in\mathds{R}^{6}$ denoting the first/second/third left (right) derivatives. As
shown in Appendix \ref{sec:remark_equivalence}, when a uniform kernel and symmetric bandwidths are used,
the constrained estimators have the same asymptotic distributions as the
unconstrained ones, thus our previous results still hold under the constrained estimates.
\end{remark}

%%%%%%%%%%%%%%%%%%%%%%%%%%%%%%%%%%%%%%%%%%%%%%%%%%%%%%%%%%%%%%%

\section{Additional Practical Considerations}

\label{sec:additional}
%%%%%%%%%%%%%%%%%%%%%%%%%%%%%%%%%%%%%%%%%%%%%%%%%%%%%%%%%%%%%%%
In order to compute the uniform consistent conditional density $f_{Y^{d}%
|VX}(\cdot|h(0),0)$ in Appendix \ref{subsec:con_den_est}, and $\mu_{1}(x,y,d)\mathbbm{1}\{|x/h_{n}%
|\leq1\}$ and $\mu_{2}(x,d)\mathbbm{1}\{|x/h_{n}|\leq1\}$ in Appendix \ref{subsec:first_stage}, we
continue to use the local cubic kernel models so the single MSE-optimal bandwidth
from the local quadratic regression can be used throughout.

\subsection{A Conditional Density Estimator}

\label{subsec:con_den_est} The statement of Theorem \ref{theorem:MB} presumes
that the densities $f_{Y^{d}|VX}(\cdot|h(0),0)$ are unknown. In order to
simulate the multiplier process, we need to replace them by their uniformly
consistent estimators. Note that the identifying formulas in Theorem
\ref{theorem:identification} suggest
\[
f_{Y^{d}|VX}(y|h(0),0)=\frac{\partial}{\partial y}F_{Y^{d}|VX}(y|h(0),0)=\frac
{\frac{\partial}{\partial y}\mu_{1}^{\prime}(0^{+},y,d)-\frac{\partial
}{\partial y}\mu_{1}^{\prime}(0^{-},y,d)}{\mu_{2}^{\prime}(0^{+},d)-\mu
_{2}^{\prime}(0^{-},d)}.
\]
Equation (\ref{eq:local_slope_estimator2}) gives uniformly consistent
estimators for the two terms in the denominator. The two terms in the
numerator can be written as
\begin{equation}
\frac{\partial}{\partial y}\mu_{1}^{\prime}(0^{\pm},y,d)=\frac{\partial
}{\partial y}\frac{\partial}{\partial x}E[\mathds{1}\{Y\leq
y\}\mathds{1}\{D=d\}|X=0^{\pm}]. \label{eq:partial_y}%
\end{equation}
With the bandwidth parameter $b_{n}$, we represent $\frac{\partial}{\partial
y}\mu_{1}(0^{\pm},y,d)$ by the limit of the regularized approximation
\begin{equation}
\mu(0^{\pm},y,d)=\lim_{n\rightarrow\infty}E\Big[\frac{1}{b_{n}}K\Big(\frac
{Y_{i}-y}{b_{n}}\Big)\mathds{1}\{D_{i}=d\}\Big|X=0^{\pm}\Big],
\label{eq:partial_y_regular}%
\end{equation}
and we estimate it by the local cubic polynomial regression
\begin{equation}
\tilde{\mu}^{\prime}(0^{\pm},y,d)a_{n}=e_{1}^{\top}%
\mathop{\rm arg~min}\limits_{\alpha\in\mathds{R}^{4}}\sum_{i=1}^{n}%
\Big[\frac{1}{b_{n}}K\Big(\frac{Y_{i}-y}{b_{n}}\Big)\mathds{1}\{D_{i}%
=d\}-r_{3}^{\top}\Big(\frac{X_{i}}{a_{n}}\Big)\alpha\Big]^{2}K\Big(\frac
{X_{i}}{a_{n}}\Big)\delta_{i}^{\pm} \label{eq:local_poly}%
\end{equation}
%\textbf{where the cubic polynomial order for the slope estimate is used to
%allow for the MSE-optimal bandwidth from the local quadratic kernel
%regression, whose bandwidth is undersmoothed for the local cubic estimand.
%}
This estimate $\tilde{\mu}^{\prime}(0^{\pm},y,d)$ is used for
(\ref{eq:partial_y}). Therefore, $\widehat{f}_{Y^{d}|VX}(y|h(0),0)$ is
estimated by
\[
\hat{f}_{Y^{d}|VX}(y|h(0),0)=\frac{\tilde{\mu}^{\prime}(0^{+},y,d)-\tilde{\mu
}^{\prime}(0^{-},y,d)}{\hat{\mu}_{2}^{\prime}(0^{+},d)-\hat{\mu}_{2}^{\prime
}(0^{-},d)}.
\]
We make the following assumption about the bandwidth parameters $a_{n}$ and
$b_{n}$.

%%%%%%%%%%%%%%%%%%%%%%%%%%%%%%%%%%%%%%%%%%%%%%

\begin{assumption}
\label{a:cond_den_est} The bandwidth parameters $a_{n}$ and $b_{n}$ satisfy
$a_{n}\rightarrow0$, $b_{n}\rightarrow0$, $na_{n}\rightarrow\infty$ and
$na_{n}^{2}b_{n}^{2}\rightarrow\infty$ and $\frac{b_n}{a_n}\to 0$.
\end{assumption}

%%%%%%%%%%%%%%%%%%%%%%%%%%%%%%%%%%%%%%%%%%%%%%
The following lemma shows that the first order derivative of the kernel
regularization (\ref{eq:partial_y_regular}) with respect to $x$ are equivalent
to the objects (\ref{eq:partial_y}) of interest. We may thus use the estimates
of $\frac{\partial}{\partial x} \mu(0^{\pm},y,d)$ to approximate
$\frac{\partial}{\partial y}\mu^{\prime}_{1}(0^{\pm},y,d)$.
%%%%%%%%%%%%%%%%%%%%%%%%%%%%%%%%%%%%%%%%%%%%%%

\begin{lemma}
\label{lemma:con_den_pop} Let Assumptions \ref{a:inference} (i) (b), (ii) (a)
(b), (iv) (a) and \ref{a:cond_den_est} hold. For each $(y,d,x)\in
\mathscr{Y}\times\mathscr{D}\times([\underline{x}, \overline x]\setminus
\{0\})$, $\frac{\partial}{\partial x} \mu(0^{\pm},y,d)=\frac{\partial
}{\partial y}\mu^{\prime}_{1}(0^{\pm},y,d)$.
\end{lemma}

%%%%%%%%%%%%%%%%%%%%%%%%%%%%%%%%%%%%%%%%%%%%%%
A proof is provided in Appendix \ref{sec:con_den_pop}. To show the uniform
consistency of $\widehat{f}_{Y^{\cdot}|VX}(\cdot|h(0),0)$, it suffices to show
$\sup_{(y,d)\in\mathscr{Y}\times\mathscr{D}}|\tilde{\mu}^{\prime}(0^{\pm
},y,d)-\mu^{\prime}(0^{\pm},y,d)|\underset{x\times\xi}{\overset{p}{\rightarrow
}}0$. The following lemma establishes this point.
%%%%%%%%%%%%%%%%%%%%%%%%%%%%%%%%%%%%%%%%%%%%%%

\begin{lemma}
\label{lemma:con_den_est_unifconst} Under Assumptions \ref{a:inference} (i),
(ii) (a) (b), (iv) (a) (b) and \ref{a:cond_den_est}, it holds that
\[
\sup_{(y,d)\in\mathscr{Y}\times\mathscr{D}}|\tilde{\mu}^{\prime}(0^{\pm
},y,d)-\mu^{\prime}(0^{\pm},y,d)|\underset{x\times\xi}{\overset{p}{\rightarrow
}}0.
\]

\end{lemma}

A proof is provided in Appendix \ref{sec:con_den_est_unifconst}.

%%%%%%%%%%%%%%%%%%%%%%%%%%%%%%%%%%%%%%%%%%%%%%

%%%%%%%%%%%%%%%%%%%%%%%%%%%%%%%%%%%%%%%%%%%%%%%%%%%%%%%%%%%%%%%

\subsection{First Stage Estimators}

\label{subsec:first_stage}
%%%%%%%%%%%%%%%%%%%%%%%%%%%%%%%%%%%%%%%%%%%%%%%%%%%%%%%%%%%%%%%
We will now give some examples of uniformly consistent estimators that satisfy
the high-level condition in Assumption \ref{a:inference} (v). First, the
density function of $X$ can be estimated by
\[
\hat f_{X}(0) = \frac{1}{n c_{n}} \sum_{i=1}^{n} K(X_{i} / c_{n}).
\]
This can be shown to be consistent if $c_{n}\to0$ and $nc_{n}\to\infty$,
$f_{X}$ is three-time differentiable and $\frac{\partial^{2}}{\partial x^{2}%
}f_{X}(0)< \infty$ -- see Theorem 1.1 of \citet{li_racine2007}.

We now propose first-stage estimators $\tilde{\mu}_{1}%
(x,y,d)\mathbbm{1}\{|x/h_{n}|\leq1\}$ and $\tilde{\mu}_{2}%
(x,d)\mathbbm{1}\{|x/h_{n}|\leq1\}$ that are used in the EMP. Denote
$\delta_{x}^{+}=\mathbbm{1}\{x\geq0\}$ and $\delta_{x}^{-}=\mathbbm{1}\{x<0\}$%
. 
We reuse the local cubic estimates from equations (\ref{eq:local_slope_estimator1}) and (\ref{eq:local_slope_estimator2}) without requiring to solve an additional optimization problem. We define the
first-stage estimators by
\begin{align*}
\tilde{\mu}_{1}(x,y,d)=  &  \Big[\widehat{\mu}_{1}(0^{+},y,d)+\widehat{\mu
}_{1}^{\prime}(0^{+},y,d)x+\widehat{\mu}_{1}^{\prime\prime}(0^{+}%
,y,d)\frac{x^{2}}{2}+\widehat{\mu}_{1}^{\prime\prime\prime}(0^{+}%
,y,d)\frac{x^{3}}{3!}\Big]\delta_{x}^{+}\\
+  &  \Big[\widehat{\mu}_{1}(0^{-},y,d)+\widehat{\mu}_{1}^{\prime}%
(0^{-},y,d)x+\widehat{\mu}_{1}^{\prime\prime}(0^{-},y,d)\frac{x^{2}}%
{2}+\widehat{\mu}_{1}^{\prime\prime\prime}(0^{-},y,d)\frac{x^{3}}%
{3!}\Big]\delta_{x}^{-}\qquad\text{and}\\
\tilde{\mu}_{2}(x,d)=  &  \Big[\widehat{\mu}_{2}(0^{+},d)+\widehat{\mu}%
_{2}^{\prime}(0^{+},d)x+\widehat{\mu}_{2}^{\prime\prime}(0^{+},d)\frac{x^{2}%
}{2}+\widehat{\mu}_{2}^{\prime\prime\prime}(0^{+},d)\frac{x^{3}}%
{3!}\Big]\delta_{x}^{+}\\
&  +\Big[\widehat{\mu}_{2}(0^{-},d)+\widehat{\mu}_{2}^{\prime}(0^{-}%
,d)x+\widehat{\mu}_{2}^{\prime\prime}(0^{-},d)\frac{x^{2}}{2}+\widehat{\mu
}_{2}^{\prime\prime\prime}(0^{-},d)\frac{x^{3}}{3!}\Big]\delta_{x}^{-}%
\end{align*}
where
\begin{align*}
&  \Big[\widehat{\mu}_{1}(0^{\pm},y,d),\widehat{\mu}_{1}^{\prime}(0^{\pm
},y,d)h_{n},\widehat{\mu}_{1}^{\prime\prime}(0^{\pm},y,d)h_{n}^{2}%
/2!,\widehat{\mu}_{1}^{\prime\prime\prime}(0^{\pm},y,d)h_{n}^{3}%
/3!\Big]^{\top}\\
&  =\mathop{\rm arg~min}\limits_{\alpha\in\mathds{R}^{4}}\sum_{i=1}%
^{n}\Big[\mathds{1}\{Y_{i}\leq y\}\mathds{1}\{D_{i}=d\}-r_{3}^{\top}%
\Big(\frac{X_{i}}{h_{n}}\Big)\alpha\Big]^{2}K\Big(\frac{X_{i}}{h_{n}%
}\Big)\delta_{i}^{\pm}\\
&  \Big[\widehat{\mu}_{2}(0^{\pm},d),\widehat{\mu}_{2}^{\prime}(0^{\pm
},d)h_{n},\widehat{\mu}_{2}^{\prime\prime}(0^{\pm},d)h_{n}^{2}/2!,\widehat{\mu
}_{2}^{\prime\prime\prime}(0^{\pm},d)h_{n}^{3}/3!\Big]^{\top}\\
&  =\mathop{\rm arg~min}\limits_{\alpha\in\mathds{R}^{4}}\sum_{i=1}%
^{n}\Big[\mathds{1}\{D_{i}=d\}-r_{3}^{\top}\Big(\frac{X_{i}}{h_{n}}%
\Big)\alpha\Big]^{2}K\Big(\frac{X_{i}}{h_{n}}\Big)\delta_{i}^{\pm}.
\end{align*}
The uniform consistency of these first-stage estimators, required as the high-level condition in Assumption \ref{a:inference} (v), follows from Lemma 7 of \citet{chiang_hsu_sasaki2019}, which is applicable under our Assumption \ref{a:inference} (i)--(iv).

%%%%%%%%%%%%%%%%%%%%%%%%%%%%%%%%%%%%%%%%%%%%%%%%%%%%%%%%%%%%%%%

\subsection{Bandwidths}

\label{subsec:bandwidth}
%%%%%%%%%%%%%%%%%%%%%%%%%%%%%%%%%%%%%%%%%%%%%%%%%%%%%%%%%%%%%%%

Another practical consideration is about a rule for selecting bandwidths in
finite sample. We propose to start with the MSE-optimal bandwidths for
local quadratic kernel smoothers as the bandwidth for
our bias-corrected local cubic kernel estimation, and then to apply the rule-of-thumb
correction for coverage optimality \citep{calonico_cattaneo_farrell2016,calonico_cattaneo_farrell2018}.
To keep the implementation simple, we use a single bandwidth $h_{n}$ that is
based on minimizing the sum of MSEs of $\overline{\mu}_{1}^{\prime}%
(0^{+},y,1)-\mu_{1}^{\prime}(0^{-},y,1)$ and $\overline{\mu}_{1}^{\prime
}(0^{+},y,0)-\mu_{1}^{\prime}(0^{-},y,0)$, where both $\overline{\mu}%
_{1}^{\prime}(0^{+},y,1)$ and $\overline{\mu}_{1}^{\prime}(0^{+},y,0)$ are
from local quadratic estimation problems. We first introduce short-hand
notations. Let $\Psi^{\pm}=\int_{\mathds{R}_{\pm}}r_{3}(u)r_{3}^{\top}%
(u)K^{2}(u)du$ and $\Lambda^{\pm}=\int_{\mathds{R}_{\pm}}u^{2}r_{3}(u)K(u)du$.

For the kernel density estimator $\hat f_{X}(0)$, we make use of Silverman's
rule of thumb
\begin{align*}
c_{n}=1.06 \hat\sigma_{X} n^{-1/5}%
\end{align*}
where $\hat\sigma_{X}$ is the sample standard deviation of $\{X_{i}%
\}^{n}_{i=1}$.

For the main bandwidth $h_{n}$, we first choose
\[
h_{0,n}=\left(  \frac{1}{2}\frac{V_{0}}{B_{0}^{2}}\right)  ^{1/7}n^{-1/7}%
\]
where the leading bias and variance terms are given by
\begin{align*}
B_{0}  &  =e_{1}^{\top}[\frac{(\Gamma^{+})^{-1}\Lambda^{+}}{3!}\overline
{\overline{\mu}}_{+}^{\prime\prime\prime}-\frac{(\Gamma^{-})^{-1}\Lambda^{-}%
}{3!}\overline{\overline{\mu}}_{-}^{\prime\prime\prime}]\qquad\text{and}\\
V_{0}  &  =\frac{e_{1}^{\top}[\overline{\bar{\sigma}}_{+}^{2}(\Gamma^{+}%
)^{-1}\Psi^{+}(\Gamma^{+})^{-1}+\overline{\bar{\sigma}}_{-}^{2}(\Gamma
^{-})^{-1}\Psi^{-}(\Gamma^{-})^{-1}]e_{1}}{\hat{f}_{X}(0)},
\end{align*}
respectively, with $\overline{\overline{\mu}}_{\pm}^{\prime\prime\prime}$ and
$\overline{\bar{\sigma}}_{\pm}^{2}$ given by global cubic parametric
regressions of $\mu_{1}^{\prime\prime\prime}(x,y,d)\delta_{x}^{\pm}$ and
$\sigma^{2}(y,d|x)\delta_{x}^{\pm}$, respectively, evaluated at $0^{\pm}$ for
certain $(y,d)$.

With the first-stage bandwidth $h_{0,n}$ having been selected, we can solve
\begin{align*}
&  \Big[\check{\mu}_{1}(0^{\pm},y,d),\check{\mu}_{1}^{\prime}(0^{\pm
},y,d)h_{0,n},\check{\mu}_{1}^{\prime\prime}(0^{\pm},y,d)h_{0,n}^{2}%
/2!,\check{\mu}_{1}^{\prime\prime\prime}(0^{\pm},y,d)h_{0,n}^{3}%
/3!\Big]^{\top}=\\
&  \qquad\qquad\mathop{\rm arg~min}\limits_{\alpha\in\mathds{R}^{4}}\sum
_{i=1}^{n}\Big[\mathds{1}\{Y_{i}\leq y\}\mathds{1}\{D_{i}=d\}-r_{3}^{\top
}\Big(\frac{X_{i}}{h_{0,n}}\Big)\alpha\Big]^{2}K(X_{i}/h_{0,n})\delta_{i}%
^{\pm},
\end{align*}
and thus compute our first-stage level estimate
\begin{align*}
\check{\mu}_{1}(x,y,d)=  &  \Big[\check{\mu}_{1}(0^{+},y,d)+\check{\mu}%
_{1}^{\prime}(0^{+},y,d)x+\check{\mu}_{1}^{\prime\prime}(0^{+},y,d)\frac
{x^{2}}{2}+\check{\mu}_{1}^{\prime\prime\prime}(0^{+},y,d)\frac{x^{3}}%
{3!}\Big]\delta_{x}^{+}\\
+  &  \Big[\check{\mu}_{1}(0^{-},y,d)+\check{\mu}_{1}^{\prime}(0^{-}%
,y,d)x+\check{\mu}_{1}^{\prime\prime}(0^{-},y,d)\frac{x^{2}}{2}+\check{\mu
}_{1}^{\prime\prime\prime}(0^{-},y,d)\frac{x^{3}}{3!}\Big]\delta_{x}^{-}.
\end{align*}

We next define the variance estimator by
\[
\hat{\sigma}(y,d|0^{\pm})=\Big(\frac{\sum_{i=1}^{n}(\mathds{1}\{Y_{i}\leq
y,D_{i}=d\}-\check{\mu}_{1}(X_{i},y,d))^{2}K(\frac{X_{i}}{h_{0,n}})\delta
_{i}^{\pm}}{\sum_{i=1}^{n}K(\frac{X_{i}}{h_{0,n}})\delta_{i}^{\pm}}\Big)^{1/2}%
\]
where $\check{\mu}_{1}(\cdot,y,d)$ is the first stage level estimator given above.

Finally, the main bandwidth selector $h_{n}$ is defined by
\[
h_{n}=\left(  \frac{1}{2}\frac{V}{B^{2}}\right)  ^{1/7}n^{-1/7}%
\]
where the leading bias and variance terms are given by
\begin{align*}
B  &  =e_{1}^{\top}[\frac{(\Gamma^{+})^{-1}\Lambda^{+}}{3!}\check{\mu}%
_{1}^{\prime\prime\prime}(0^{+},y,d)-\frac{(\Gamma^{-})^{-1}\Lambda^{-}}%
{3!}\check{\mu}_{1}^{\prime\prime\prime}(0^{-},y,d)]\qquad\text{and}\\
V  &  =\frac{e_{1}^{\top}[\hat{\sigma}(y,d|0^{+})(\Gamma^{+})^{-1}\Psi
^{+}(\Gamma^{+})^{-1}+\hat{\sigma}(y,d|0^{-})(\Gamma^{-})^{-1}\Psi^{-}%
(\Gamma^{-})^{-1}]e_{1}}{\hat{f}_{X}(0)}.
\end{align*}
In the
end, following \citet{calonico_cattaneo_farrell2016,calonico_cattaneo_farrell2018}, we can apply the
rule-of-thumb (ROT) correction for coverage optimality bandwidth of the
local quadratic regression to the main bandwidth as $h^{ROT}_n=n^{-2/35}h_{n}.$

For the bandwidth parameters $a_{n}$ and $b_{n}$ used for the conditional density estimator $\widehat{f}_{Y^{d}|VX}(y|h(0),0)$ in Appendix \ref{subsec:con_den_est}, we follow the choice rules proposed in the end of Appendix C in \citet{frandsen_frolich_melly2012}, and propose to set $a_{n}=h_{n}$ and $b_{n}=h^2_{n}%
$.

%%%%%%%%%%%%%%%%%%%%%%%%%%%%%%%%%%%%%%%%%%%%%%%%%%%%%%%%%%%%%%%%%%%%%%%%%%%%%%%%%%%%%%%%%%%%%%%

\section{Auxiliary Lemmas and Proofs}
\label{sec:auxiliary_lemmas_and_proofs}
%%%%%%%%%%%%%%%%%%%%%%%%%%%%%%%%%%%%%%%%%%%%%%%%%%%%%%%%%%%%%%%%%%%%%%%%%%%%%%%%%%%%%%%%%%%%%%%

\subsection{Auxiliary Lemmas}

%%%%%%%%%%%%%%%%%%%%%%%%%%%%%%%%%%%%%%%%%%%%%%%%%%%%%%%%%%%%%%%%%%%%%%%%%%%%%%%%%%%%%%%%%%%%%%%

\subsubsection{Uniform Bahadur Representation}

%%%%%%%%%%%%%%%%%%%%%%%%%%%%%%%%%%%%%%%%%%%%%%%%%%%%%%%%%%%%%%%%%%%%%%%%%%%%%%%%%%%%%%%%%%%%%%%
The following lemma proposes the uniform BR for the local slope estimators.

\begin{lemma}
[\citet{chiang_hsu_sasaki2019}; Lemma 1]\label{lemma:br} Under Assumption
\ref{a:inference}, we have the uniform influence function representations
(\ref{eq:br1}) and (\ref{eq:br2}) that hold uniformly on $\mathscr{Y}_{1}%
\times\mathscr{D}$.
\end{lemma}

%%%%%%%%%%%%%%%%%%%%%%%%%%%%%%%%%%%%%%%%%%%%%%%%%%%%%%%%%%%%%%%%%%%%%%%%%%%%%%%%%%%%%%%%%%%%%%%

\subsubsection{Functional Central Limit Theorem}

%%%%%%%%%%%%%%%%%%%%%%%%%%%%%%%%%%%%%%%%%%%%%%%%%%%%%%%%%%%%%%%%%%%%%%%%%%%%%%%%%%%%%%%%%%%%%%%

\begin{lemma}
\label{lemma:FCLT} Let triangular array of separable stochastic processes
$\{f_{ni}(\omega,t):i=1,...n,t\in T\}$ be row independent and write
$X_{n}(t)=\sum_{i=1}^{n}[f_{ni}(\omega,t)-Ef_{ni}(\cdot,t)]$, and denote
$E^{\ast}$ to be the outer integral (see, e.g., Section 1.2 of van der Vaart
and Wellner (1996)). Suppose that the following conditions are satisfied:

\begin{enumerate}
\item $\left\{  f_{ni}\right\}  $ are manageable, with envelope $\left\{
F_{ni}\right\}  $ which are also independent within rows;

\item $H(s,t)=\lim_{n\rightarrow\infty}EX_{n}(s)X_{n}(t)$ exists for every
$s,t\in T$;

\item $\limsup_{n\to\infty} \sum_{i=1}^{n}E^{*}F^{2}_{ni}<\infty$;

\item $\lim_{n\to\infty} \sum_{i=1}^{n}E^{*}F^{2}_{ni}\mathbbm{1}\{F_{ni}%
>\epsilon\}=0$ for each $\epsilon>0$;

\item $\rho(s,t)=\lim_{n\rightarrow\infty}\rho_{n}(s,t),$ where $\rho
_{n}(s,t)=(\sum_{i=1}^{n}E[f_{ni}(\cdot,s)-f_{ni}(\cdot,t)]^{2})^{1/2},$
exists for every $s,t\in T$, and for all deterministic sequences $\{s_{n}\}$
and $\{t_{n}\}$ in $\mathds{T}$, if $\rho(s_{n},t_{n})\rightarrow0$ then
$\rho_{n}(s_{n},t_{n})\rightarrow0$.
\end{enumerate}

Then $T$ is totally bounded under the $\rho$ pseudometric, and $X_{n}$
converges weakly to a tight mean zero Gaussian process $\mathds{X}$
concentrated on $\left\{  z\in l^{\infty}\left(  T\right)  :z\text{ is
uniformly }\rho-\text{continuous}\right\}  $, with covariance $H(s,t)$.
\end{lemma}

%%%%%%%%%%%%%%%%%%%%%%%%%%%%%%%%%%%%%%%%%%%%%%%%%%%%%%%%%%%%%%%%%%%%%%%%%%%%%%%%%%%%

%%%%%%%%%%%%%%%%%%%%%%%%%%%%%%%%%%%%%%%%%%%%%%%%%%%%%%%%%%%%%%%%%%%%%%%%%%%%%%%%%%%%

\subsection{Proof of Theorem \ref{theorem:MB}}

\label{sec:MB}

Before starting to present a proof of the theorem, we introduce additional
definitions and notations for the proof of the theorem. Let $\mathcal{F}$ be a
class of measurable functions defined on $(\Omega,\mathscr{F})$ with a
measurable envelope $F$. We say that $\mathcal{F}$ is of VC type with envelope
$F$ if there exist constants $A$, $v>0$ such that $\sup_{Q} N(\mathcal{F}%
,L^{2}(Q),\varepsilon\left\|  F\right\|  _{Q,2})$ $\le$ $(A/\varepsilon)^{v}$,
where the supremum is taken over the set of all finite discrete measures $Q$
on $\mathcal{F}$.

To approximate the distribution of the BR, we define the following Multiplier
Processes (MP):
\begin{align}
\nu_{\xi,n}^{\pm}(y,d,1)=  &  \frac{1}{\sqrt{nh_{n}}f_{X}(0)}\sum_{i=1}^{n}%
\xi_{i}e_{1}^{\top}(\Gamma^{\pm})^{-1}r_{3}\Big(\frac{X_{i}}{h_{n}%
}\Big)\Big[\mathbbm{1}\{Y_{i}\leq y,D_{i}=d\}-\mu_{1}(X_{i}%
,y,d)\Big]K\Big(\frac{X_{i}}{h_{n}}\Big)\delta_{i}^{\pm},\nonumber\\
\nu_{\xi,n}^{\pm}(y,d,2)=  &  \frac{1}{\sqrt{nh_{n}}f_{X}(0)}\sum_{i=1}^{n}%
\xi_{i}e_{1}^{\top}(\Gamma^{\pm})^{-1}r_{3}\Big(\frac{X_{i}}{h_{n}%
}\Big)\Big[\mathbbm{1}\{D_{i}=d\}-\mu_{2}(X_{i},d)\Big]K\Big(\frac{X_{i}%
}{h_{n}}\Big)\delta_{i}^{\pm}.\nonumber
\end{align}
For ease of writing, we use the following notations for the differences of
right and left limits of the BR, the MP, and the EMP with $k=1,2$:
\begin{align*}
\nu_{n}(y,d,k)  &  =\nu_{n}^{+}(y,d,k)-\nu_{n}^{-}(y,d,k),\\
\nu_{\xi,n}(y,d,k)  &  =\nu_{\xi,n}^{+}(y,d,k)-\nu_{\xi,n}^{-}(y,d,k),\\
\hat{\nu}_{\xi,n}(y,d,k)  &  =\hat{\nu}_{\xi,n}^{+}(y,d,k)-\hat{\nu}_{\xi
,n}^{-}(y,d,k).
\end{align*}
With these preparations, we now start a proof of Theorem \ref{theorem:MB}.

\textbf{Part (i) (a):} We will verify the five conditions in Lemma
\ref{lemma:FCLT} for the triangular array of stochastic processes $\{f_{ni}\}$
defined by
\begin{align*}
f_{ni}(y,d,1)=  &  \frac{1}{\sqrt{nh_{n}}f_{X}(0)}e_{1}^{\top}(\Gamma
^{+})^{-1}r_{3}\Big(\frac{X_{i}}{h_{n}}\Big)\Big[\mathbbm{1}\{Y_{i}\leq
y\}\mathbbm{1}\{D_{i}=d\}-\mu_{1}(X_{i},y,d)\Big]K\Big(\frac{X_{i}}{h_{n}%
}\Big)\delta_{i}^{+},\\
f_{ni}(y,d,2)=  &  \frac{1}{\sqrt{nh_{n}}f_{X}(0)}e_{1}^{\top}(\Gamma
^{+})^{-1}r_{3}\Big(\frac{X_{i}}{h_{n}}\Big)\Big[\mathbbm{1}\{D_{i}%
=d\}-\mu_{2}(X_{i},y,d)\Big]K\Big(\frac{X_{i}}{h_{n}}\Big)\delta_{i}^{+},\\
\nu_{n}^{+}(y,d,k)=  &  \sum_{i=1}^{n}[f_{ni}(y,d,k)-Ef_{ni}(y,d,k)].
\end{align*}
The separability follows the same argument as in the proof of Theorem 4 of
\citet{kosorok2003} and the left or right continuity of the processes. To show
condition 1, define
\begin{align*}
\mathscr{F}_{n}  &  =\{(y^{\ast},d^{\ast},x^{\ast})\mapsto\mathbbm{1}\{x^{\ast
}\geq0\}[(\mathbbm{1}\{y^{\ast}\leq y,d^{\ast}=d\}-\mu_{1}(x^{\ast
},y,d))\mathbbm{1}\{k=1\}\\
&  \quad+(\mathbbm{1}\{d^{\ast}=d\}-\mu_{2}(x^{\ast}%
,y,d))\mathbbm{1}\{k=2\}]:(y,d,k)\in\mathscr{Y}_{1}\times\{0,1\}\times
\{1,2\}\}\\
\mathscr{F}_{n}^{+}  &  =\{f_{ni}(y,d,k):(y,d,k)\in\mathscr{Y}_{1}%
\times\{0,1\}\times\{1,2\}\}
\end{align*}
We first claim that $\mathscr{F}_{n}^{+}$ is a VC type class with envelope
$$
F_{n}^{+}(y^{\ast},d^{\ast},x^{\ast})=\frac{C^{\prime\prime}}{\sqrt{nh_{n}}}\left\Vert K\right\Vert _{\infty}\mathbbm{1}\{|x^{\ast}/h_{n}|\in \lbrack-1,1]\}
$$ 
for some constant $C^{\prime\prime}>0$. It is clear
$\{(y^{\ast},d^{\ast},x^{\ast})\mapsto\mathbbm{1}\{y^{\ast}\leq y\}:y\in
\mathscr{Y}_{1}\}$ is of VC-subgraph with VC index $\leq2,$ since it is
monotone increasing in $y$, and thus for each pair $(y_{1}^{\ast},x_{1}^{\ast
},d_{1}^{\ast},r_{1}),(y_{2}^{\ast},x_{2}^{\ast},d_{2}^{\ast},r_{2}%
)\in\mathscr{Y}_{1}\times\mathscr{X}\times\{0,1\}\times\mathds{R}$ with
$y_{1}^{\ast}\leq y_{2}^{\ast}$, it can never pick out $\{(y_{2}^{\ast}%
,x_{2}^{\ast},d_{2}^{\ast},r_{2})\}$. Similarly, $\{(y^{\ast},d^{\ast}%
,x^{\ast})\mapsto\mathbbm{1}\{d^{\ast}=d\}:d\in\{1,2\}\}$, $\{(y^{\ast
},d^{\ast},x^{\ast})\mapsto\{\mathbbm{1}\{k^{\ast}=k\}:k\in\{1,2\}\}$ and
$\{(y^{\ast},d^{\ast},x^{\ast})\mapsto\mathbbm{1}\{x^{\ast}\geq0\}\}$ are all
VC subgraph classes, since they are sub-collections of all half spaces and
then by Lemma 9.12 (i) of \citet{kosorok2008}. Each of them is therefore of VC type
with envelope $1$. Next, Assumption \ref{a:inference}(ii)(a)(b) imply
$$
|\mu_{k_{1}}(x^{\ast},y_{1},d_{1})-\mu_{k_{2}}(x^{\ast},y_{2},d_{2})|\leq
L\left\Vert (k_{1},y_{1},d_{1})-(k_{2},y_{2},d_{2})\right\Vert 
$$ 
for an $L>0$
and Euclidean norm $\left\Vert \cdot\right\Vert $. Thus $\{x^{\ast}\mapsto
\mu_{k}(x,y,d):(k,y,d)\in\{1,2\}\times\mathscr{Y}_{1}\times\mathscr{D}\}$ is
of VC type with envelope $L$ in light of Example 19.7 of van der Vaart (1998)
and Lemma 9.18 of \citet{kosorok2008}. Under Assumption \ref{a:inference}(i)(b),
(iii) and (iv), for each $n$, the collection of a single function 
$$
\{(y^{\ast
},d^{\ast},x^{\ast})\mapsto\frac{e_{1}^{\top}(\Gamma^{+})^{-1}r_{3}(x^{\ast
}/h_{n})\mathbbm{1}\{|x^{\ast}/h_{n}|\in\lbrack-1,1]\}}{\sqrt{nh_{n}}f_{X}%
(0)}\}
$$ 
is of VC subgraph and therefore VC type with envelope $\frac
{C^{\prime}\mathbbm{1}\{|x^{\ast}/h_{n}|\in\lbrack-1,1]\}}{\sqrt{nh_{n}}}$.
Example 19.19 of van der Vaart (1998) suggests VC type classes, that are of
finite uniform integrals, are closed under element-wise addition and
multiplication. Therefore, $\mathscr{F}_{n}$ is of VC type with envelope
constant $C^{\prime\prime}$ and thus
\[
\mathscr{F}_{n}^{+}=\{\frac{e_{1}^{\top}(\Gamma^{+})^{-1}r_{3}(\cdot
/h_{n})K(\cdot/h_{n})}{\sqrt{nh_{n}}f_{X}(0)}\cdot f:f\in\mathscr{F}_{n}\}
\]
is of VC type with envelope $F_{n}^{+}(y^{\ast},d^{\ast},x^{\ast}%
)=\frac{C^{\prime\prime}}{\sqrt{nh_{n}}}\left\Vert K\right\Vert _{\infty
}\mathbbm{1}\{x^{\ast}/h_{n}\in\lbrack-1,1]\}$. Finally, standard calculations
show for each $n$ and for any $\delta\in(0,1)$ the uniform entropy integral
bound
\[
\int_{0}^{\delta}\sup_{Q}\sqrt{1+\log N(\mathscr{F}_{n},L_{2}(Q),\varepsilon
\left\Vert F_{n}\right\Vert _{Q,2})}d\varepsilon\lesssim\delta\sqrt
{v\log(A/\delta)}.
\]
Equation (A.1) in the proof of Theorem 1 in Andrews (1994) then implies that
$\mathscr{F}_{n}^{+}$ is a manageable class of functions, as defined in
Section 11.4.1 of \citet{kosorok2008}. To check condition 2, notice 
\begin{align*}
E\nu_{n}%
^{+}(y_{1},d_{1},k_{1})\nu_{n}^{+}(y_{2},d_{2},k_{2})=\sum_{i=1}^{n}%
Ef_{ni}(y_{1},d_{1},k_{1})f_{ni}(y_{2},d_{2},k_{2})
\\
-(\sum_{i=1}^{n}%
Ef_{ni}(y_{1},d_{1},k_{1}))(\sum_{i=1}^{n}Ef_{ni}(y_{2},d_{2},k_{2})). 
\end{align*}
It
suffices to check $\sum_{i=1}^{n}Ef_{ni}(y_{1},d_{1},k_{1})f_{ni}(y_{2}%
,d_{2},k_{2})<\infty$ since $Ef_{ni}(y,d,k)=0$ due to the law of iterated
expectations, and thus the second term is $0$. When $k_{1}=k_{2}=1$, under
Assumption \ref{a:inference}(i)(a)(b),(ii)(c),(iii), (iv)(a),
\begin{align*}
&  \sum_{i=1}^{n}Ef_{ni}(y_{1},d_{1},1)f_{ni}(y_{2},d_{2},1)\\
=  &  E[\frac{e_{1}^{\top}(\Gamma^{+})^{-1}r_{3}(\frac{X_{i}}{h_{n}}%
)r_{3}^{\top}(\frac{X_{i}}{h_{n}})(\Gamma^{+})^{-1}e_{1}}{h_{n}f_{X}^{2}%
(0)}[\mathbbm{1}\{Y_{i}\leq y_{1},D_{i}=d_{1}\}-\mu_{1}(X_{i},y_{1},d_{1})]\\
&  \qquad\times\lbrack\mathbbm{1}\{Y_{i}\leq y_{2},D_{i}=d_{2}\}-\mu_{1}%
(X_{i},y_{2},d_{2})]K^{2}(\frac{X_{i}}{h_{n}})\delta_{i}^{+}]\\
=  &  \int_{\mathds{R}_{+}}\frac{e_{1}^{\top}(\Gamma^{+})^{-1}r_{3}%
(u)r_{3}^{\top}\left(  u\right)  (\Gamma^{+})^{-1}e_{1}}{f_{X}^{2}(0)}%
K^{2}(u)\left(  \sigma_{11}((y_{1},d_{1}),(y_{2},d_{2})|0^{+})+O\left(
uh_{n}\right)  \right)  (f_{X}(0)+O(uh_{n}))du\\
=  &  \int_{\mathds{R}_{+}}\frac{e_{1}^{\prime}(\Gamma^{+})^{-1}%
r(u)r^{\prime+})^{-1}e_{1}}{f_{X}(0)}K^{2}(u)\sigma_{11}((y_{1},d_{1}%
),(y_{2},d_{2})|0^{+})du+O(h_{n})<\infty
\end{align*}
where the second to the last equality is due to mean value expansions under
Assumption \ref{a:inference} (i)(b) and (ii)(c). Notice that $n$ enters only
through the $O(h_{n})$ term, and thus 
$$
\lim_{n\rightarrow\infty}\sum_{i=1}%
^{n}Ef_{ni}(y_{1},d_{1},1)f_{ni}(y_{2},d_{2},1)
$$ 
exists. Similar calculations
hold for $k_{1}=k_{2}=1$ and $k_{1}=1$, $k_{2}=2$. This shows condition 2.
Condition 3 is clear since 
\begin{align*}
\lim_{n\rightarrow\infty}\sum_{i=1}^{n}E\left[
F_{n}^{+}(y^{\ast},d^{\ast},x^{\ast})\right]  ^{2}=\lim_{n\rightarrow\infty
}\int\frac{(C^{\prime\prime2}}{h_{n}}\left\Vert K\right\Vert _{\infty}%
^{2}\mathbbm{1}\{|x/h_{n}|\in\lbrack-1,1]\}f_{X}(x)dx
\\
=f(0)(C^{\prime\prime
2}\left\Vert K\right\Vert _{\infty}^{2}<\infty
\end{align*}
under Assumption
\ref{a:inference} (i)(a), (iii) and (iv)(a). To show condition 4, note that
for each $\varepsilon>0$,
\begin{align*}
&  \lim_{n\rightarrow\infty}\sum_{i=1}^{n}E[\left(  F_{n}^{+}(y^{\ast}%
,d^{\ast},x^{\ast})\right)  ^{2}\mathbbm{1}\{F_{n}^{+}(y^{\ast},d^{\ast
},x^{\ast})>\varepsilon\}]\\
=  &  \lim_{n\rightarrow\infty}\int_{\mathds{R}}\frac{(C^{\prime\prime2}%
}{h_{n}}\left\Vert K\right\Vert _{\infty}^{2}\mathbbm{1}\{x/h_{n}\in
\lbrack-1,1]\}\mathbbm{1}\{\frac{(C^{\prime\prime})}{nh_{n}}\left\Vert
K\right\Vert _{\infty}\mathbbm{1}\{x/h_{n}\in\lbrack-1,1]\}>\varepsilon
\}f_{X}(x)dx\\
\leq &  \int_{\mathds{R}}(C^{\prime\prime2}\left\Vert K\right\Vert _{\infty
}^{2}\mathbbm{1}\{u\in\lbrack-1,1]\}\lim_{n\rightarrow\infty}%
\mathbbm{1}\{\frac{(C^{\prime\prime})}{nh_{n}}\left\Vert K\right\Vert
_{\infty}\mathbbm{1}\{u\in\lbrack-1,1]\}>\varepsilon\}f_{X}(0)du+O(h_{n})=0
\end{align*}
under Assumption \ref{a:inference} (i)(a), (iii) and (iv)(a). This shows
condition 4. To show condition 5, note that we can write
\begin{align*}
\rho_{n}^{2}((y_{1},d_{1},k_{1}),(y_{2},d_{2},k_{2}))  &  =\sum_{i=1}%
^{n}E[f_{ni}(y_{1},d_{1},k_{1})-f_{ni}(y_{2},d_{2},k_{2})]^{2}\\
&  =\sum_{i=1}^{n}E[f_{ni}^{2}(y_{1},d_{1},k_{1})+f_{ni}^{2}(y_{2},d_{2}%
,k_{2})-2f_{ni}(y_{1},d_{1},k_{1})f_{ni}(y_{2},d_{2},k_{2})].
\end{align*}
From our calculations on the way to show condition 2, we know that each term
on the right-hand side exists under Assumption \ref{a:inference}
(i)(a)(b),(ii)(c),(iii), (iv)(a). Since $n$ enters the expression only through
the $O(h_{n})$ part, for all deterministic sequences $s_{n}\in\mathscr{Y}_{1}%
\times\{0,1\}\times\{1,2\}$ and $t_{n}\in\mathscr{Y}_{1}\times\{0,1\}\times
\{1,2\}$, $\rho^{2}(s_{n},t_{n})\rightarrow0$ implies $\rho_{n}^{2}%
(s_{n},t_{n})\rightarrow0.$ By Lemma 4, we have $\nu_{n}^{+}\leadsto
\mathds{G}_{+}$ and similarly for $\nu_{n}^{-}\leadsto\mathds{G}_{-}$.
Assumption \ref{a:inference}(i)(a) then implies $\nu_{n}=\nu_{n}^{+}-\nu
_{n}^{-}\leadsto\mathds{G}:=\mathds{G}_{+}-\mathds{G}_{-}$. \newline%
\textbf{Part (i) (b):} We apply the FCLT and the functional delta method.
Notice that $\nu_{n}\leadsto\mathds{G}$ suggests
\[
\sqrt{nh_{n}^{3}}%
\begin{bmatrix}
(\hat{\mu}_{1}^{\prime}(0^{+},y,d)-\hat{\mu}_{1}^{\prime}(0^{-},y,d))-(\mu
_{1}^{\prime}(0^{+},y,d)-\mu_{1}^{\prime}(0^{-},y,d))\\
(\hat{\mu}_{2}^{\prime}(0^{+},d)-\hat{\mu}_{2}^{\prime}(0^{-},d))-(\mu
_{2}^{\prime}(0^{+},d)-\mu_{2}^{\prime}(0^{-},d))
\end{bmatrix}
=%
\begin{bmatrix}
\mathds{G}(y,d,1)\\
\mathds{G}(y,d,2)
\end{bmatrix}
.
\]
Let $(A(\cdot),B(\cdot))\in\ell^{\infty}(\mathscr{Y}_{1}\times\{0,1\})\times
\ell^{\infty}(\mathscr{Y}_{1})$, if $B(\cdot)>C>0$, then $(G,H)\overset{\Psi
}{\mapsto}{G}/{H}$ is Hadamard differentiable at $(A,B)$ tangentially to
$\ell^{\infty}$ with the Hadamard derivative $\Psi_{(A,B)}^{\prime}$ given by
$\Psi_{(A,B)}^{\prime}(g,h)={(Bg-Ah)}/{B^{2}}$. Therefore, under Assumption
\ref{a:rkd}(ii), we know that $\mu_{2}^{\prime}(0^{+},d)-\mu_{2}^{\prime
}(0^{-},d)$ is bounded away from $0$. Also, $f_{Y^{d}|VX}(\cdot|h(0),0)$ is
bounded away from zero under Assumption \ref{a:inference}(i)(c). The
functional delta method then yields
\begin{align*}
&  \sqrt{nh_{n}^{3}}[\widehat{F}_{Y^{d}|VX}(\cdot|h(0),0)-F_{Y^{d}|VX}%
(\cdot|h(0),0)]\\
=  &  \sqrt{nh_{n}^{3}}[\frac{\hat{\mu}_{1}^{\prime}(0^{+},\cdot,d)-\hat{\mu
}_{1}^{\prime}(0^{-},\cdot,d)}{\hat{\mu}_{2}^{\prime}(0^{+},d)-\hat{\mu}%
_{2}^{\prime}(0^{-},d)}-\frac{\mu_{1}^{\prime}(0^{+},\cdot,d)-\mu_{1}^{\prime
}(0^{-},\cdot,d)}{\mu_{2}^{\prime}(0^{+},d)-\mu_{2}^{\prime}(0^{-},d)}]\\
\leadsto &  \mathds{G}_{F}(\cdot,d)
\end{align*}
where
\[
\mathds{G}_{F}(y,d):=\frac{[\mu_{2}^{\prime}(0^{+},d)-\mu_{2}^{\prime}%
(0^{-},d)]\mathds{G}(y,d,1)-[\mu_{1}^{\prime}(0^{+},y,d)-\mu_{1}^{\prime
}(0^{-},y,d)]\mathds{G}(y,d,2)}{[\mu_{2}^{\prime}(0^{+},d)-\mu_{2}^{\prime
}(0^{-},d)]^{2}}.
\]
\newline\textbf{Part (i) (c):} Define operator $\Upsilon:\mathds{D}_{\Upsilon
}(\mathscr{Y}_{1}\times\{0,1\})\rightarrow\ell^{\infty}([a,1-a])$ as
\[
F(\cdot,\cdot)\overset{\Upsilon}{\mapsto}\Phi(F(\cdot,1))\left(  \cdot\right)
-\Phi(F(\cdot,0))(\cdot)=Q(\cdot,1)-Q(\cdot,0)
\]
where $\Phi(F)(\theta)=Q(\theta)=\inf\{y\in\mathscr{Y}_{1}:F(y)\geq\theta\}$.
By Hadamard differentiability from Lemma 3.9.23(ii) of van der Vaart and
Wellner (1996) and the chain rule (van der Vaart, 1998, Theorem 20.9),
under Assumption \ref{a:inference}(i)(c),(ii)(a)(b), $\Upsilon$ is Hadamard
differentiable at $F_{Y^{\cdot}|VX}(\cdot|h(0),0)$ tangentially to
$\mathcal{C}(\mathscr{Y}_{1}\times\mathscr{D})$ and the derivative
\citep[][Section 2.2.4]{kosorok2008} is%
\begin{align*}
&  \Upsilon_{F_{Y^{\cdot}|VX}(\cdot|h(0),0)}^{\prime}(g(\cdot,\cdot))\\
=  &  -\frac{g(Q_{Y^{1}|VX}(\cdot|h(0),0),1)}{f_{Y^{1}|VX}(Q_{Y^{1}|VX}%
(\cdot|h(0),0)|h(0),0)}+\frac{g(Q_{Y^{0}|VX}(\cdot|h(0),0),0)}{f_{Y^{0}%
|VX}(Q_{Y^{0}|VX}(\cdot|h(0),0)|h(0),0)}%
\end{align*}
is tangential to $C(\mathscr{Y}_{1}\times\mathscr{D})$. The functional delta
method then yields
\[
\sqrt{nh_{n}^{3}}[\hat{\tau}-\tau]\leadsto\mathds{G}_{\tau}%
\]
where
\[
\mathds{G}_{\tau}(\theta)=-\Big[\frac{\mathds{G}_{F}(Q_{Y^{1}|VX}%
(\theta|h(0),0),1)}{f_{Y^{1}|VX}(Q_{Y^{1}|VX}(\theta|h(0),0)|h(0),0)}%
-\frac{\mathds{G}_{F}(Q_{Y^{0}|VX}(\theta|h(0),0),1)}{f_{Y^{0}|VX}%
(Q_{Y^{0}|VX}(\theta|h(0),0)|h(0),0)}\Big].
\]
\newline\textbf{Part (ii):} This part of the proof consists of two steps. We
first show the convergence result for the EMP, and then show the convergence
result for $\widehat{\Xi}\left(  \cdot\right)  $.\newline\textbf{Step 1} We
claim $\nu_{\widehat{\xi},n}\underset{\xi}{\overset{p}{\leadsto}}\mathds{G}$.
Applying Theorem 11.19 of \citet{kosorok2008}, which is applicable under the five
conditions verified in (i), we have $\nu_{\xi,n}=\nu_{\xi,n}^{+}-\nu_{\xi
,n}^{-}\underset{\xi}{\overset{p}{\leadsto}}\mathds{G}$. In light of of Lemma
2 of \citet{chiang_hsu_sasaki2019}, it then suffices to show
\[
\underset{(y,d,k)\in\mathscr{Y}_{1}\times\{0,1\}\times\{1,2\}}{\sup}|\hat{\nu
}_{\xi,n}^{\pm}(y,d,k)-\nu_{\xi,n}^{\pm}(y,d,k)|\underset{x\times
\xi}{\overset{p}{\rightarrow}}0.
\]
Indeed, for $k=1$, by Assumption \ref{a:inference}(i)(b),(v), we have
\begin{align}
&  \hat{\nu}_{\xi,n}^{+}(y,d,1)-\nu_{\xi,n}^{+}(y,d,1)\nonumber\\
=  &  \frac{1}{f_{X}(0)\hat{f}_{X}(0)}\sum_{i=1}^{n}\xi_{i}\frac{e_{1}^{\top
}\left(  \Gamma^{+}\right)  ^{-1}r_{3}(\frac{X_{i}}{h_{n}})K(\frac{X_{i}%
}{h_{n}})\delta_{i}^{+}}{\sqrt{nh_{n}}}[\mathds{1}\{Y_{i}\leq y,D_{i}%
=d\}f_{X}(0)-\tilde{\mu}_{1}(0^{+},y,d)f_{X}(0)\nonumber\\
&  -\mathds{1}\{Y_{i}\leq y,D_{i}=d\}\hat{f}_{X}(0)+\mu_{1}(0^{+},y,d)\hat
{f}_{X}(0)]\nonumber\\
=  &  \frac{1}{f_{X}^{2}(0)+o_{p}^{x\times\xi}(1)}\sum_{i=1}^{n}T_{i}%
^{+}[-\tilde{\mu}_{1}(0^{+},y,d)f_{X}(0)+\mu_{1}(0^{+},y,d)\hat{f}%
_{X}(0)+o_{p}^{x\times\xi}(1)]\nonumber\\
=  &  \frac{1}{f_{X}^{2}(0)+o_{p}^{x\times\xi}(1)}\sum_{i=1}^{n}T_{i}%
^{+}[-\tilde{\mu}_{1}(0^{+},y,d)f_{X}(0)+\mu_{1}(0^{+},y,d)f_{X}(0)\nonumber\\
&  -\mu_{1}(0^{+},y,d)f_{X}(0)+\mu_{1}(0^{+},y,d)\hat{f}_{X}(0)+o_{p}%
^{x\times\xi}(1)]\nonumber\\
=  &  \frac{f_{X}(0)}{f_{X}^{2}(0)+o_{p}^{x\times\xi}(1)}\sum_{i=1}^{n} T_{i}^{+}[-\tilde{\mu}_{1}(0^{+},y,d)+\mu_{1}(0^{+},y,d)]\nonumber\\
&  +\frac{\mu_{1} (0^{+},y,d)}{f_{X}^{2}(0)+o_{p}^{x\times\xi}(1)}\sum_{i=1}^{n}T_{i}^{+} [-f_{X}(0)+\hat{f}_{X}(0)]+\frac{o_{p}^{x\times\xi}(1)}{f_{X}^{2}(0)+o_{p}^{x\times\xi}(1)}\sum
_{i=1}^{n}T_{i}^{+}\nonumber\\
=  &  \frac{f_{X}(0)}{f_{X}^{2}(0)+o_{p}^{x\times\xi}(1)}\sum_{i=1}^{n} T_{i}^{+}o_{p}^{x\times\xi}(1)+\frac{\mu_{1}(0^{+},y,d)}{f_{X}^{2} (0)+o_{p}^{x\times\xi}(1)}\sum_{i=1}^{n}T_{i}^{+}o_{p}^{x\times\xi }(1)\nonumber\\
&  +\frac{o_{p}^{x\times\xi}(1)}{f_{X}^{2}(0)+o_{p}^{x\times\xi}(1)}\sum
_{i=1}^{n}T_{i}^{+}\nonumber\\
=  &  o_{p}^{x\times\xi}(1)\sum_{i=1}^{n}T_{i}^{+} \label{eq:est_errors}%
\end{align}
where $T_{i}^{+}=\xi_{i}\frac{e_{1}^{\top}\left(  \Gamma^{+}\right)
^{-1}r_{3}(\frac{X_{i}}{h_{n}})K(\frac{X_{i}}{h_{n}})\delta_{i}^{+}}%
{\sqrt{nh_{n}}}$. It can be shown that the array of zero mean random variables
$\{\sum_{i=1}^{n}T_{i}^{+}\}_{i=1}^{n}$ satisfies Lindeberg-Feller conditions
(Proposition 2.27 of van der Vaart (1998)) under Assumption \ref{a:inference}%
(i)(a), (iii) and (iv)(a)(c) and therefore converges in distribution to a
normal distribution. Therefore, the asymptotic tightness then implies
$\sum_{i=1}^{n}T_{i}=O_{p}^{x\times\xi}(1)$. Thus we conclude that equation
\ref{eq:est_errors} is $o_{p}^{x\times\xi}(1)$. \qquad\newline\textbf{Step 2}
We will show
\[
-[\frac{\hat{Z}_{\xi,n}(\hat{Q}_{Y^{1}|VX}(\cdot|h(0),0),1)}{\hat{f}%
_{Y^{1}|VX}(\hat{Q}_{Y^{1}|VX}(\cdot)|h(0),0)}-\frac{\hat{Z}_{\xi,n}(\hat
{Q}_{Y^{0}|VX}(\cdot|h(0),0),0)}{\hat{f}_{Y^{0}|VX}(\hat{Q}_{Y^{0}|VX}%
(\cdot)|h(0),0)}]\underset{\xi}{\overset{p}{\leadsto}}\mathds{G}_{\tau}%
(\cdot)
\]
where
\[
\hat{Z}_{\xi,n}(y,d)=\frac{[\hat{\mu}_{2}^{\prime}(0^{+},d)-\hat{\mu}%
_{2}^{\prime}(0^{-},d)]\hat{\nu}_{\xi,n}(y,d,1)-[\hat{\mu}_{1}^{\prime}%
(0^{+},y,d)-\hat{\mu}_{1}^{\prime}(0^{-},y,d)]\hat{\nu}_{\xi,n}(y,d,2)}%
{[\hat{\mu}_{2}^{\prime}(0^{+},d)-\hat{\mu}_{2}^{\prime}(0^{-},d)]^{2}}.
\]
We first use Theorem 12.1 of \citet{kosorok2008} (the functional delta for
bootstrap) along with the conclusion of Step 1 to get
\[
\tilde{Z}_{\xi,n}(\cdot,\cdot):=\frac{[\mu_{2}^{\prime}(0^{+},\cdot)-\mu
_{2}^{\prime}(0^{-},\cdot)]\hat{\nu}_{\xi,n}(\cdot,\cdot,1)-[\mu_{1}^{\prime
}(0^{+},\cdot,\cdot)-\mu_{1}^{\prime}(0^{-},\cdot,\cdot)]\hat{\nu}_{\xi
,n}(\cdot,\cdot,2)}{[\mu_{2}^{\prime}(0^{+},\cdot)-\mu_{2}^{\prime}%
(0^{-},\cdot)]^{2}}\underset{\xi}{\overset{p}{\leadsto}}\mathds{G}_{F}%
(\cdot,\cdot).
\]
Since the denominator is bounded away from $0$ under Assumption
\ref{a:inference}(i)(iv), uniform consistency of $\hat{\mu}_{1}^{^{\prime}}$,
$\hat{\mu}_{2}^{^{\prime}}$ from Theorem 2 gives $\left\Vert \tilde{Z}_{\xi
,n}-\hat{Z}_{\xi,n}\right\Vert _{\mathscr{Y}_{1}\times\{0,1\}}%
\underset{x\times\xi}{\overset{p}{\rightarrow}}0$, and Lemma 2 of \citet{chiang_hsu_sasaki2019} implies $\hat{Z}_{\xi,n}\underset{\xi}{\overset{p}{\leadsto
}}\mathds{G}_{F}$. Using the functional delta method for bootstrap again, we
obtain
\[
-[\frac{\hat{Z}_{\xi,n}(Q_{Y^{1}|VX}(\cdot|h(0),0),1)}{f_{Y^{1}|VX}%
(Q_{Y^{1}|VX}(\cdot|h(0),0)|h(0),0)}-\frac{\hat{Z}_{\xi,n}(Q_{Y^{0}|VX}%
(\cdot|h(0),0),0)}{f_{Y^{0}|VX}(Q_{Y^{0}|VX}(\cdot|h(0),0)|h(0),0)}%
]\underset{\xi}{\overset{p}{\leadsto}}\mathds{G}_{\tau}(\cdot).
\]
Since $f_{Y^{d}|VX}(\cdot|h(0),0)$ are bounded away from zero, using
asymptotic $\rho-$equicontinuity of $\hat{Z}_{\xi,n}(\cdot,\cdot)$ following
its (conditional) weak convergence and Theorem 3.7.23 of \citet{gine_nickl2016}, and the uniform consistency of $\hat{f}_{Y^{d}|VX}(\cdot|h(0),0)$ and
$\hat{Q}_{Y^{d}|VX}(\cdot)$ with $d=1,2$ along with Lemma 2 of \citet{chiang_hsu_sasaki2019}, we conclude part (ii) of the theorem.
%%%%%%%%%%%%%%%%%%%%%%%%%%%%%%%%%%%%%%%%%%%%%%%%%%%%%%%%%%%%%%%%%%%%%%%%%%%%%%%%%%%%

%%%%%%%%%%%%%%%%%%%%%%%%%%%%%%%%%%%%%%%%%%%%%%%%%%%%%%%%%%%%%%%%%%%%%%%%%%%%%%%%%%%%
%\subsection{Derivation of Equation (\ref{eq:first_stage_estimand})}\label{sec:first_stage_estimand}

%Given the absolute continuity of the conditional distribution of $Y^d$ given $(V,X)$, the identifying formulae in Theorem \ref{theorem:identification} allow us to write
%\begin{align}
%f_{Y^d|VX}(y|h(0),0)=\frac{\partial}{\partial y}F_{Y^d|VX}(y|h(0),0)&=\frac{\frac{\partial}{\partial y}\mu'_1(0^+,y,d)-\frac{\partial}{\partial y}\mu'_1(0^-,y,d)}{\mu'_2(0^+,d)-\mu'_2(0^-,d)}.\nonumber
%\end{align}
%The terms in the numerator can be written as
%\begin{align*}
%\frac{\partial}{\partial y}\mu'_1(0^\pm,y,d)&=P_{D|X}(d|0^\pm)\frac{\partial}{\partial x}f_{Y|DX}(y|d,0^\pm)+f_{Y|DX}(y|d,0^\pm)\frac{\partial}{\partial x}P_{D|X}(d|0^\pm)
%\end{align*}
%We have $P_{D|X}(d|0^+)=P_{D|X}(d|0^-)=P_{D|X}(d|0)$ and $f_{Y|DX}(y|d,0^+)=f_{Y|DX}(y|d,0^-)=f_{Y|DX}(y|d,0)$ under Assumption.
%Thus we have
%\begin{align*}
%&f_{Y^d|VX}(y|h(0),0)\\
%=&\frac{P_{D|X}(d|0)[\frac{\partial}{\partial x}f_{Y|DX}(y|d,0^+)-\frac{\partial}{\partial x}f_{Y|DX}(y|d,0^-)]+f_{Y|DX}(y|d,0)[\frac{\partial}{\partial x}P_{D|X}(d|0^+)-\frac{\partial}{\partial x}P_{D|X}(d|0^-)]}{\mu'_2(0^+,d)-\mu'_2(0^-,d)}
%\end{align*}
%as desired.
%\qed
%%%%%%%%%%%%%%%%%%%%%%%%%%%%%%%%%%%%%%%%%%%%%%%%%%%%%%%%%%%%%%%%%%%%%%%%%%%%%%%%%%%%

\subsection{Proof of Lemma \ref{lemma:con_den_pop}}

\label{sec:con_den_pop} We prove the lemma by two steps: for each
$(y,d,x)\in\mathscr{Y}\times\mathscr{D}\times([\underline{x},\overline
{x}]\setminus\{0\})$, Step 1 shows 
$$\frac{\partial}{\partial x}\mu
(0^{\pm},y,d)=\frac{\partial}{\partial x}(f_{Y|XD}(y|x,d)P_{D|X}(d|x))
$$
and
Step 2 shows 
$$
\frac{\partial}{\partial y}\mu_{1}^{\prime}(0^{\pm}%
,y,d)=\frac{\partial}{\partial y}\frac{\partial}{\partial x}%
E[\mathds{1}\{Y_{i}\leq y,D_{i}=d\}|X_{i}=x]=\frac{\partial}{\partial
x}(f_{Y|XD}(y|x,d)P_{D|X}(d|x)).
$$

\textbf{Step 1} For $d=1$, under Assumptions \ref{a:inference} (i) (b), (ii)
(a) (b), (iv) (a) and \ref{a:cond_den_est}, for each $(y,x)\in
\mathscr{Y}\times([\underline{x},\overline{x}]\setminus\{0\})$, for $d=1$,
applying the dominated convergence theorem, we have
\begin{align*}
&  \frac{\partial}{\partial x}\lim_{n\rightarrow\infty}E\Big[\frac{1}{b_{n}%
}K\Big(\frac{Y_{i}-y}{b_{n}}\Big)\mathds{1}\{D_{i}=1\}|X_{i}=x\Big]\\
=  &  \frac{\partial}{\partial x}\lim_{n\rightarrow\infty}(E\Big[\frac
{1}{b_{n}}K\Big(\frac{Y_{i}-y}{b_{n}}\Big)\Big|X_{i}=x\Big]P_{D|X}(1|x)+0)\\
=  &  \frac{\partial}{\partial x}\lim_{n\rightarrow\infty}\Big(\int%
_{\mathds{R}}K(u)f_{Y|XD}(ub_{n}+y|x,1)duP_{D|X}(1|x)\Big)\\
=  &  \frac{\partial}{\partial x}\lim_{n\rightarrow\infty}\Big(\int%
_{\mathds{R}}K(u)(f_{Y|XD}(y|x,1)+\frac{\partial}{\partial y}f_{Y|XD}%
(y|x,1)ub_{n}+\frac{\partial^{2}}{\partial y^{2}}f_{Y|XD}(y^{\ast}%
|x,1)\frac{u^{2}b_{n}^{2}}{2})duP_{D|X}(1|x)\Big)\\
=  &  \frac{\partial}{\partial x}\lim_{n\rightarrow\infty}((f_{Y|XD}%
(y|x,1)+O(b_{n}^{2}))P_{D|X}(1|x))=\frac{\partial}{\partial x}(f_{Y|XD}%
(y|x,1)P_{D|X}(1|x)).
\end{align*}
where $y\ast$ lies between $y$ and $y+ub_{n}$. Similar result holds for $d=0$.

\textbf{Step 2} Under Assumptions \ref{a:inference} (i) (b), (ii) (a) (b),
(iv) (a) and \ref{a:cond_den_est}, for each $(y,x)\in\mathscr{Y}\times
([\underline{x},\overline{x}]\setminus\{0\})$, for $d=1$, an application of
the dominated convergence theorem yields
\begin{align*}
\frac{\partial}{\partial y}\frac{\partial}{\partial x}E\Big[\mathds{1}\{Y_{i}%
\leq y,D_{i}=1\}\Big|X_{i}=x\Big]=  &  \frac{\partial}{\partial y}%
\frac{\partial}{\partial x}\Big(E\Big[\mathds{1}\{Y_{i}\leq y\}\Big|X_{i}%
=x\Big]P_{D|X}(1|x)+0\Big)\\
=  &  \frac{\partial}{\partial y}\frac{\partial}{\partial x}F_{Y|XD}%
(y|x,1)P_{D|X}(1|x)\\
=  &  \frac{\partial}{\partial x}\frac{\partial}{\partial y}F_{Y|XD}%
(y|x,1)P_{D|X}(1|x)=\frac{\partial}{\partial x}f_{Y|XD}(y|x,1)P_{D|X}(1|x).
\end{align*}
Similar result holds for $d=0$. \qed

%%%%%%%%%%%%%%%%%%%%%%%%%%%%%%%%%%%%%%%%%%%%%%%%%%%%%%%%%%%%%%%%%%%%%%%%%%%%%%%%%%%%%

\subsection{Proof of Lemma \ref{lemma:con_den_est_unifconst}}

\label{sec:con_den_est_unifconst} The proof makes use of a maximal
inequality from \citet{chernozhukov_chetverikov_kato2014}. Under Assumptions
\ref{a:inference} (ii) (a) (b) and \ref{a:cond_den_est}, as in Section 1.6 of
\citet{tsybakov2008}, the solution to equation (\ref{eq:local_poly}) can be written
as
\begin{align*}
&  \tilde{\alpha}(0^{+},y,d)\\
=  &  \Big[\frac{1}{na_{n}}\sum_{i=1}^{n}\delta_{i}^{+}K\Big(\frac{X_{i}%
}{a_{n}}\Big)r_{3}\Big(\frac{X_{i}}{a_{n}}\Big)r_{3}^{\top}\Big(\frac{X_{i}%
}{a_{n}}\Big)\Big]^{-1}\Big[\frac{1}{na_{n}}\sum_{i=1}^{n}\delta_{i}%
^{+}K\Big(\frac{X_{i}}{a_{n}}\Big)r_{3}\Big(\frac{X_{i}}{a_{n}}\Big)\Big(\frac
{1}{b_{n}}K\Big(\frac{Y_{i}-y}{b_{n}}\Big)\mathds{1}\{D_{i}=d\}\Big)\Big]\\
=  &  \alpha(0^{+},y,d)+\Big[\frac{1}{na_{n}}\sum_{i=1}^{n}\delta_{i}%
^{+}K\Big(\frac{X_{i}}{a_{n}}\Big)r_{3}\Big(\frac{X_{i}}{a_{n}}\Big)r_{3}%
^{\top}\Big(\frac{X_{i}}{a_{n}}\Big)\Big]^{-1}\frac{1}{na_{n}}\sum_{i=1}%
^{n}\delta_{i}^{+}K\Big(\frac{X_{i}}{a_{n}}\Big)r_{3}\Big(\frac{X_{i}}{a_{n}%
}\Big)\frac{\mu^{(4)}(x_{ni}^{\ast},y,d)}{4!}a_{n}^{4}\\
+  &  \Big[\frac{1}{na_{n}}\sum_{i=1}^{n}\delta_{i}^{+}K\Big(\frac{X_{i}%
}{a_{n}}\Big)r_{3}\Big(\frac{X_{i}}{a_{n}}\Big)r_{3}^{\top}\Big(\frac{X_{i}%
}{a_{n}}\Big)\Big]^{-1}\\
&  \Big[\frac{1}{na_{n}}\sum_{i=1}^{n}\delta_{i}^{+}K\Big(\frac{X_{i}}{a_{n}%
}\Big)r_{3}\Big(\frac{X_{i}}{a_{n}}\Big)\Big(\frac{1}{b_{n}}K\Big(\frac
{Y_{i}-y}{b_{n}}\Big)\mathds{1}\{D_{i}=d\}-\mu(X_{i},y,d)\Big)\Big]
\end{align*}
where $\alpha(0^{+},y,d)=\Big[\mu(0^{\pm},y,d),\mu^{\prime}(0^{\pm}%
,y,d)a_{n},\mu^{\prime\prime}(0^{\pm},y,d)a_{n}^{2}/2!,\mu^{\prime\prime
\prime}(0^{\pm},y,d)a_{n}^{3}/3!\Big]^{\top}$. Multiply both sides by
$e_{1}^{\top}$ to get
\[
\tilde{\mu}^{\prime}(0^{+},y,d)=\mu^{\prime}(0^{+},y,d)+(1)+(2)
\]
where
\begin{align*}
(1)=  &  e_{1}^{\top}\Big[\frac{1}{na_{n}}\sum_{i=1}^{n}\delta_{i}%
^{+}K\Big(\frac{X_{i}}{a_{n}}\Big)r_{3}\Big(\frac{X_{i}}{a_{n}}\Big)r_{3}%
^{\top}\Big(\frac{X_{i}}{a_{n}}\Big)\Big]^{-1}\frac{1}{na_{n}}\sum_{i=1}%
^{n}\delta_{i}^{+}K\Big(\frac{X_{i}}{a_{n}}\Big)r_{3}\Big(\frac{X_{i}}{a_{n}%
}\Big)\frac{\mu^{(4)}(x_{ni}^{\ast},y,d)}{4!}a_{n}^{4}\\
(2)=  &  e_{1}^{\top}\Big[\frac{1}{na_{n}}\sum_{i=1}^{n}\delta_{i}%
^{+}K\Big(\frac{X_{i}}{a_{n}}\Big)r_{3}\Big(\frac{X_{i}}{a_{n}}\Big)r_{3}%
^{\top}\Big(\frac{X_{i}}{a_{n}}\Big)\Big]^{-1}\\
&  \frac{1}{na_{n}}\sum_{i=1}^{n}\delta_{i}^{+}K\Big(\frac{X_{i}}{a_{n}%
}\Big)r_{3}\Big(\frac{X_{i}}{a_{n}}\Big)\Big(\frac{1}{b_{n}}K\Big(\frac
{Y_{i}-y}{b_{n}}\Big)\mathds{1}\{D_{i}=d\}-\mu(X_{i},y,d)\Big).
\end{align*}
From Step 1 of Proof of Lemma 1 in \citet{chiang_hsu_sasaki2019}, with
Assumption \ref{a:inference} (i) (a) (b), (iii) and (iv) and
\ref{a:cond_den_est}, we have the common inverse factor
\[
\Big[\frac{1}{na_{n}}\sum_{i=1}^{n}\delta_{i}^{+}K\Big(\frac{X_{i}}{a_{n}%
}\Big)r_{3}\Big(\frac{X_{i}}{a_{n}}\Big)r_{3}^{\top}\Big(\frac{X_{i}}{a_{n}%
}\Big)\Big]^{-1}\underset{x\times\xi}{\overset{p}{\rightarrow}}\frac
{(\Gamma^{+})^{-1}}{f_{X}(0)}%
\]
uniformly in $(y,d)$. It suffices to show that each of
\begin{align*}
(3)=  &  \frac{1}{na_{n}}\sum_{i=1}^{n}\delta_{i}^{+}K\Big(\frac{X_{i}}{a_{n}%
}\Big)r_{3}\Big(\frac{X_{i}}{a_{n}}\Big)\frac{\mu^{(4)
}(x_{ni}^{\ast},y,d)}{4!}a_{n}^{4}\\
(4)=  &  \frac{1}{na_{n}}\sum_{i=1}^{n}\delta_{i}^{+}K\Big(\frac{X_{i}}{a_{n}%
}\Big)r_{3}\Big(\frac{X_{i}}{a_{n}}\Big)\Big(\frac{1}{b_{n}}K\Big(\frac
{Y_{i}-y}{b_{n}}\Big)\mathds{1}\{D_{i}=d\}-\mu(X_{i},y,d)\Big)
\end{align*}
converges in probability to zero uniformly. We will divide the argument into
the following four steps.

\noindent\textbf{Step 1} Under Assumption \ref{a:inference} (i)(a),
(ii)(a)(b), (iii) and (iv)(a), it holds that
\begin{align*}
\Big|\frac{1}{na_{n}}\sum_{i=1}^{n}\delta_{i}^{+}K\Big(\frac{X_{i}}{a_{n}%
}\Big)r_{3}\Big(\frac{X_{i}}{a_{n}}\Big)\frac{\mu^{(4)
}(x_{ni}^{\ast},y,d)}{4!}a_{n}^{4}\Big|\leq &  \frac{1}{na_{n}}\sum_{i=1}%
^{n}\Big|K\Big(\frac{X_{i}}{a_{n}}\Big)\Big|\Big|r_{3}\Big(\frac{X_{i}}{a_{n}%
}\Big)\Big|\Big|\frac{\mu^{(4)}(x_{ni}^{\ast},y,d)}%
{4!}a_{n}^{4}\Big|\\
\lesssim &  \frac{n}{na_{n}}\left\Vert K\right\Vert _{\infty}Ma_{n}%
^{4}\rightarrow0.
\end{align*}

\noindent\textbf{Step 2} We first bound the difference
\begin{align*}
&  \frac{1}{na_{n}b_{n}}\sum_{i=1}^{n}\delta_{i}^{+}K\Big(\frac{X_{i}}{a_{n}%
}\Big)r_{3}\Big(\frac{X_{i}}{a_{n}}\Big)K\Big(\frac{Y_{i}-y}{b_{n}%
}\Big)\mathds{1}\{D_{i}=d\}\\
&  -E\Big[\frac{1}{na_{n}b_{n}}\sum_{i=1}^{n}\delta_{i}^{+}K\Big(\frac{X_{i}%
}{a_{n}}\Big)r_{3}\Big(\frac{X_{i}}{a_{n}}\Big)K\Big(\frac{Y_{i}-y}{b_{n}%
}\Big)\mathds{1}\{D_{i}=d\}\Big].
\end{align*}
It suffices to show that each term converges in probability uniformly. Define
for each $t=0,1,...,3$
\begin{align*}
&  \mathscr{F}_{t}=\{(y^{\ast},d^{\ast},x^{\ast})\longmapsto\delta_{x}%
^{+}(ax^{\ast})^{t}K(ax^{\ast})K(by^{\ast}+c)\mathds\{d^{\ast}=d\}:d\in
\mathscr{D},a,b\geq0,c\in\mathds{R}\}\qquad\text{and}\\
&  \mathscr{F}_{t,n}=\{(y^{\ast},d^{\ast},x^{\ast})\longmapsto\delta_{x}%
^{+}(x/a_{n})^{t}K(x/a_{n})K((y^{\ast}-y)/b_{n})\mathds\{d^{\ast}%
=d\}:d\in\mathscr{D},y\in\mathscr{Y}_{1}\}.
\end{align*}
where $\delta_{x}^{+}=\mathbbm{1}\{x\geq0\}$ and $\delta_{x}^{-}%
=\mathbbm{1}\{x<0\}$. Note that for a fixed $t$, $\mathscr{F}_{t,n}%
\subset\mathscr{F}_{t}$ for all $n$. Fix any $t$, under Assumption
\ref{a:inference} (iv), $\{x^{\ast}\mapsto K(ax^{\ast}):a\in\mathds{R}\}$ is
of VC Type class with measurable envelope $\left\Vert K\right\Vert _{\infty}$.
By Proposition 3.6.12 of \citet{gine_nickl2016}, $x\mapsto(ax)^{t}%
\mathds{1}\{ax\leq1\}$ is of VC type class with measurable envelope $1$ since
$z\mapsto z^{t}\mathds{1}\{z\leq1\}$ is a mapping of bounded variations.
Furthermore, $\{\mathds{1}\{d^{\ast}=d\}:d\in\mathscr{D}\}$ is of VC-subgraph
class and therefore of VC type. Lemma A.6 of \citet{chernozhukov_chetverikov_kato2014} then implies that the class of their element-wise product
$\mathscr{F}_{t}$ is of VC type with envelope $F_{t}=\left\Vert K\right\Vert
_{\infty}^{2}$, i.e., there exist positive constants $k$, $v<\infty$ such that
$\sup_{Q}N(\mathscr{F}_{t},\left\Vert \cdot\right\Vert _{Q,2},\varepsilon
\left\Vert F_{t}\right\Vert _{Q,2})\leq(\frac{k}{\varepsilon})^{v}$ for
$0<\varepsilon\leq1$ and the supremum is taken over the set of all probability
measures on $(\Omega^{x},\mathcal{F}^{x})$. Corollary 5.1 in \citet{chernozhukov_chetverikov_kato2014} then gives
\[
E\Bigg[\left\Vert \frac{1}{\sqrt{n}}\sum_{i=1}^{n}(f(Y_{i},D_{i}%
,X_{i})-Ef(Y_{i},D_{i},X_{i}))\right\Vert _{\mathscr{F}_{t}}\Bigg]=O_{p}%
^{x}(1).
\]
Multiplying both sides by $(\sqrt{n}a_{n}b_{n})^{-1}$, we have
\begin{align*}
E\Bigg[\underset{(y,d)\in\mathscr{Y}_{1}\times\mathscr{D}}{\sup}  &
\Bigg|\frac{1}{na_{n}b_{n}}\sum_{i=1}^{n}\delta_{i}^{+}K\Big(\frac{X_{i}%
}{a_{n}}\Big)r_{3}\Big(\frac{X_{i}}{a_{n}}\Big)K\Big(\frac{Y_{i}-y}{b_{n}%
}\Big)\mathds{1}\{D_{i}=d\}\\
&  -E\Big[\frac{1}{na_{n}b_{n}}\sum_{i=1}^{n}\delta_{i}^{+}K\Big(\frac{X_{i}%
}{a_{n}}\Big)r_{3}\Big(\frac{X_{i}}{a_{n}}\Big)K\Big(\frac{Y_{i}-y}{b_{n}%
}\Big)\mathds{1}\{D_{i}=d\}\Big]\Bigg|\Bigg]=O(\frac{1}{\sqrt{n}a_{n}b_{n}}).
\end{align*}
The result then follows from Markov's inequality and Assumption
\ref{a:cond_den_est}.

\noindent\textbf{Step 3} We now want to control
\[
\frac{1}{na_{n}}\sum_{i=1}^{n}\delta_{i}^{+}K\Big(\frac{X_{i}}{a_{n}%
}\Big)r_{3}\Big(\frac{X_{i}}{a_{n}}\Big)\mu(X_{i},y,d)-E\Big[\frac{1}{na_{n}%
}\sum_{i=1}^{n}\delta_{i}^{+}K\Big(\frac{X_{i}}{a_{n}}\Big)r_{3}%
\Big(\frac{X_{i}}{a_{n}}\Big)\mu(X_{i},y,d)\Big].
\]
Since under Assumption \ref{a:inference} (ii)(a)(b), for any $(y_{1},d_{1})$,
$(y_{2},d_{2})\in\mathscr{Y}\times\mathscr{D}$, $|\mu(x,y_{1},d_{1}%
)-\mu(x,y_{2},d_{2})|\leq M(x)(|y_{1}-y_{2}|+|d_{1}-d_{2}|)$, this implies
that $\{\mu(\cdot,y,d):y\in\mathscr{Y}_{1},d\in\mathscr{D}\}$ is of VC type
class in lieu of Example 19.7 of van der Vaart (1998) and Lemma 9.18 of
\citet{kosorok2008}. We can then follow the same steps as in Step 2 to show
\begin{align*}
&  E\Bigg[\underset{(y,d)\in\mathscr{Y}_{1}\times\mathscr{D}}{\sup}%
\Bigg|\frac{1}{na_{n}}\sum_{i=1}^{n}\delta_{i}^{+}K\Big(\frac{X_{i}}{a_{n}%
}\Big)r_{3}\Big(\frac{X_{i}}{a_{n}}\Big)\mu(X_{i},y,d)-E\Big[\frac{1}{na_{n}%
}\sum_{i=1}^{n}\delta_{i}^{+}K\Big(\frac{X_{i}}{a_{n}}\Big)r_{3}%
\Big(\frac{X_{i}}{a_{n}}\Big)\mu(X_{i},y,d)\Big]\Bigg|\Bigg]\\
&  =O\Big(\frac{1}{\sqrt{n}a_{n}}\Big).
\end{align*}
The desired result of the current step then follows from Markov's inequality
and Assumption \ref{a:cond_den_est}.

\noindent\textbf{Step 4} Finally, we show that the two expectations above are
asymptotically equivalent uniformly in $y$ and $d$. Under Assumption
\ref{a:inference} (i) (b), (ii) (a) (b), (iii), (iv) (a), calculations yield
\begin{align*}
&  E\Big[\frac{1}{na_{n}b_{n}}\sum_{i=1}^{n}\delta_{i}^{+}K\Big(\frac{X_{i}%
}{a_{n}}\Big)r_{3}\Big(\frac{X_{i}}{a_{n}}\Big)K\Big(\frac{Y_{i}-y}{b_{n}%
}\Big)\mathds{1}\{D_{i}=d\}\Big]=\\
&  E\Big[\frac{1}{na_{n}}\sum_{i=1}^{n}\delta_{i}^{+}K\Big(\frac{X_{i}}{a_{n}%
}\Big)r_{3}\Big(\frac{X_{i}}{a_{n}}\Big)\mu(X_{i},y,d)\Big]
\end{align*}
by the law of iterated expectations under Assumption \ref{a:cond_den_est}.
This result, along with results from Steps 2 and 3, concludes the proof.
\qed

%%%%%%%%%%%%%%%%%%%%%%%%%%%%%%%%%%%%%%%%%%%%%%%%%%%%%%%%%%%%%%%%%%%%%%%%%%%%%%%%%%%%%

\subsection{On Remark \ref{remark:equivalence}}\label{sec:remark_equivalence}

This appendix section proves the statement in Remark \ref{remark:equivalence}.
We mostly follow the proof of Proposition 6 of \citet{card_lee_pei_weber2015}.
Let $\mathds{1}\{\mathbf{Y}\leq y\}\mathds{1}\{\mathbf{D}=d\}$ be the
\textquotedblleft stacked\textquotedblright\ $n\times1$ outcome variable
$\left\{  \mathds{1}\{Y_{i}\leq y\}\mathds{1}\{D_{i}=d\}\right\}  _{i=1}^{n}$,
where the first $n^{-}$ entries are observations to the left of $x_{0}$ and
the last $n^{+}$ entries are those to the right of $x_{0}$. Let $\mathbf{Z}$
be the $n\times8$ matrix whose $i^{\text{th}}$ row is%
\begin{align*}
\left(  \delta^-_i,\frac{X_{i}}{h_{n}}\delta^-_i,\left(  \frac{X_{i}}{h_{n}}\right)
^{2}\delta^-_i,\left(  \frac{X_{i}}{h_{n}}\right)  ^{3}\delta^-_i,\delta^+_i,\frac{X_{i}}{h_{n}}\delta^+_i,\left(  \frac{X_{i}}{h_{n}%
}\right)^{2}\delta^+_i,\left(  \frac{X_{i}}{h_{n}}\right)^{3}\delta^+_i\right).
\end{align*}
Also let 
$$
\mathbf{W}_{K}=\left(
\begin{array}
[c]{cc}%
\mathbf{W}_{K}^{-} & 0\\
0 & \mathbf{W}_{K}^{+}%
\end{array}
\right)  
$$ 
with $\mathbf{W}_{K}^{\pm}$ being the diagonal matrices
$$
\mathbf{Diag}\left(  K\Big(\frac{X_{1}}{h_{n}}\Big)\delta^\pm_1,....,K\Big(\frac
{X_{n}}{h_{n}}\Big)\delta^\pm_n\right) .
%=\mathbf{Diag}\left(  K\Big(\frac{X_{1}^{\pm}}{h_{n}}\Big),...., %\Big(\frac
%{X_{n}^{\pm}}{h_{n}}\Big)\right)  
$$
The constrained estimator can be obtained
from
\[
\min_{\beta^{R}\in\mathcal{R}^{8}}\left(  \mathds{1}\{\mathbf{Y}\leq
y\}\mathds{1}\{\mathbf{D}=d\}-\mathbf{Z}\beta^{R}\right)  ^{\top}%
\mathbf{W}_{K}\left(  \mathds{1}\{\mathbf{Y}\leq y\}\mathds{1}\{\mathbf{D}%
=d\}-\mathbf{Z}\beta^{R}\right)
\]
subject to $\mathbf{R}\beta^{R}=0$ where $\mathbf{R=}\left(
1,0,0,0,-1,0,0,0\right)  $. Denote the resulting estimator by
\[
\widehat{\beta}^{R}=\left(
\begin{array}
[c]{c}%
\hat{\mu}_{1}^{R}(0^{+},y,d),\hat{\mu}_{1}^{\prime R}(0^{+},y,d)h_{n},\hat
{\mu}_{1}^{\prime\prime R}(0^{+},y,d)h_{n}^{2}/2!,\hat{\mu}_{1}^{\prime
\prime\prime R}(0^{+},y,d)h_{n}^{3}/3!,\\
\hat{\mu}_{1}^{R}(0^{-},y,d),\hat{\mu}_{1}^{\prime R}(0^{-},y,d)h_{n},\hat
{\mu}_{1}^{\prime\prime R}(0^{-},y,d)h_{n}^{2}/2!,\hat{\mu}_{1}^{\prime
\prime\prime R}(0^{-},y,d)h_{n}^{3}/3!
\end{array}
\right)  .
\]
From equation (1.4.5) of \citet{amemiya1985}, we have%
\begin{align*}
&  \widehat{\beta}^{R}-\beta\\
&  =\left[
\begin{array}
[c]{c}%
\left(  \mathbf{Z}^{\top}\mathbf{W}_{K}\mathbf{Z}\right)  ^{-1}\\
-\left(  \mathbf{Z}^{\top}\mathbf{W}_{K}\mathbf{Z}\right)  ^{-1}%
\mathbf{R}^{\top}\left(  \mathbf{R}\left(  \mathbf{Z}^{\top}\mathbf{W}%
_{K}\mathbf{Z}\right)  ^{-1}\mathbf{R}^{\top}\right)  ^{-1}\mathbf{R}\left(
\mathbf{Z}^{\top}\mathbf{W}_{K}\mathbf{Z}\right)  ^{-1}%
\end{array}
\right]  \mathbf{Z}^{\top}\mathbf{W}_{K}\left(  \mathds{1}\{\mathbf{Y}\leq
y\}\mathds{1}\{\mathbf{D}=d\}-\mathbf{Z}\beta\right)  \\
&  =\left(  \mathbf{Z}^{\top}\mathbf{W}_{K}\mathbf{Z}\right)  ^{-1}%
\mathbf{Z}^{\top}\mathbf{W}_{K}\left(  \mathds{1}\{\mathbf{Y}\leq
y\}\mathds{1}\{\mathbf{D}=d\}-\mathbf{Z}\beta\right)  \\
&  -\left(  \mathbf{Z}^{\top}\mathbf{W}_{K}\mathbf{Z}\right)  ^{-1}%
\mathbf{R}^{\top}\left(  \mathbf{R}\left(  \mathbf{Z}^{\top}\mathbf{W}%
_{K}\mathbf{Z}\right)  ^{-1}\mathbf{R}^{\top}\right)  ^{-1}\mathbf{R}\left(
\mathbf{Z}^{\top}\mathbf{W}_{K}\mathbf{Z}\right)  ^{-1}\cdot\mathbf{Z}^{\top
}\mathbf{W}_{K}\left(  \mathds{1}\{\mathbf{Y}\leq y\}\mathds{1}\{\mathbf{D}%
=d\}-\mathbf{Z}\beta\right)  \\
&  =\left(  \mathbf{Z}^{\top}\mathbf{W}_{K}\mathbf{Z}\right)  ^{-1}%
\mathbf{Z}^{\top}\mathbf{W}_{K}\left(  \mathds{1}\{\mathbf{Y}\leq
y\}\mathds{1}\{\mathbf{D}=d\}-\mathbf{Z}\beta\right)  \\
&  -\Pi^{-1}\mathbf{R}^{\top}\left(  \mathbf{R}\Pi\mathbf{^{-1}R}^{\top
}\right)  ^{-1}\mathbf{R}\Pi^{-1}\cdot\mathbf{Z}^{\top}\mathbf{W}_{K}\left(
\mathds{1}\{\mathbf{Y}\leq y\}\mathds{1}\{\mathbf{D}=d\}-\mathbf{Z}%
\beta\right)  \frac{1}{nh_{n}}+o_{p}\left(  \frac{1}{nh_{n}}\right),
\end{align*}
where the first term on the RHS is the unconstrained version and $\Pi^{-1}$ is
$$
\Pi
^{-1}=\left(
\begin{array}
[c]{cc}%
\Gamma^{-} & 0\\
0 & \Gamma^{+}%
\end{array}
\right).
$$
Since $\hat{\mu}_{1}^{\prime R}(0^{+},y,d)h_{n}-\hat{\mu}%
_{1}^{\prime R}(0^{-},y,d)h_{n}=\mathbf{E}\widehat{\beta}^{R}$, where
$\mathbf{E=}\left(  0,1,0,0,0,-1,0,0\right)  $ and $K$ is the uniform kernel,
we have $\mathbf{E\cdot}\Pi^{-1}\cdot\mathbf{R}^{\top}=0$. Therefore,
\[
\hat{\mu}_{1}^{\prime R}(0^{+},y,d)h_{n}-\hat{\mu}_{1}^{\prime R}%
(0^{-},y,d)h_{n}=\mathbf{E\cdot}\left(  \mathbf{Z}^{\top}\mathbf{W}%
_{K}\mathbf{Z}\right)  ^{-1}\mathbf{Z}^{\top}\mathbf{W}_{K}\left(
\mathds{1}\{\mathbf{Y}\leq y\}\mathds{1}\{\mathbf{D}=d\}-\mathbf{Z}%
\beta\right)  +o_{p}\left(  \frac{1}{nh_{n}}\right),
\]
where the constrained estimator has the same asymptotic distribution as the
unconstrained one. \qed

\end{document}